\documentclass[aps,prb,preprint,groupedaddress]{revtex4-1}
\usepackage{chemformula} 
\usepackage[T1]{fontenc} 
\usepackage{amsmath}
\usepackage{amssymb}
\usepackage{epsfig}
\usepackage{graphicx}
\usepackage{braket}
\usepackage{url}
\usepackage{subcaption}
\usepackage{booktabs}
\usepackage{multirow}

\begin{document}


\title{Pressure induced electronic structure transformation of topological semimetal}


\author{Feihu Liu}
\email[]{liufeihu@xupt.edu.cn}
\affiliation{School of Science, XI'AN University of Posts and Telecommunications, XI'AN, China}
\author{Lina Wu}
\email{wulina02@gmail.com}
\affiliation{School of Science, XI'AN Technological University, XI'AN, China}


\date{\today}

\begin{abstract}
We study the electronic structure change of yttrium trihydride \ch{YH_3} by applying a hydrostatic pressure. At zero pressure, \ch{YH_3} has the structure with energy favored symmetry group $p\bar{3}c1$ (165). From first principle calculation, we argue that the band crossings are caused by overlapping of an electron- and hole-like bands. Besides the space inversion symmetry ($\mathcal{I}$) and the time reversal symmetry, the band structure also exhibits an approximate particle-hole symmetry.
Thus, \ch{YH_3} can be viewed as a pseudo nodal surface semimetal belongs to class BDI of the ten-fold AZ+ $\mathcal{I}$ classifications of gapless topological matter \cite{bzdusek_robust_2017}. As pressure increases, the approximate particle-hole symmetry is gradually broken and the pseudo nodal surface turns into a nodal ring belonging to the class AI with fewer non-spatial symmetries. Also, the nodal ring is shrinking in the process. At about $31$ GPa, which is higher than the reported structure phase transition pressure $21$ GPa, the nodal ring shrinks to a nodal  point. When above $31$ GPa, all band crossings are gapped out and \ch{YH_3} becomes a trivial insulator eventually. 
\end{abstract}


\maketitle


The topological phase of matter has became the most attractive subject of modern condensed matter physics since the discovery of the topological insulators \cite{fu_topological_2007,hasan_colloquium_2010}. A lots of progress has been made to understand the fully gapped topological phases. Rather remarkably, many of the interesting parts can be understand via free fermion theory \cite{witten_three_2016}. In fact, for the gapped free fermion systems, the topological phases protected by the non-spatial symmetries, namely, time reversal symmetry ($\mathcal{T}$), particle-hole symmetry ($\mathcal{P}$) and chiral symmetry ($\mathcal{C}$), can be classified in a ten-fold way \cite{schnyder_classification_2008,kitaev_periodic_2009,ryu_topological_2010}. This classification sheme can be naturally generalized for gapless topological matters, i.e, topological semimetal \cite{nature_topological_2016,burkov_topological_2016,xu_discovery_2015,lu_experimental_2015,lv_experimental_2015}. For such systems, the band touching is oftenly protected by both crystalline symmetries and non-spatial symmetries. Due to the fact that the non-spatial symmetries are more stable compared to crystalline symmetries, at first it seems a good idea to search for the nodes protected by non-spatial symmetries only. However the result is quite disappointing because there are very few of them. In order to enlarge the set of stable nodes, it was proposed by Ref. \onlinecite{bzdusek_robust_2017,fang_topological_2015} that one could consider space inversion symmetry ($\mathcal{I}$) too. Since space inversion symmetry is more stable compared to rotations and reflections (for example, against straining), one can obtain a similar ten-fold $\mathrm{AZ}+\mathcal{I}$ classification of the so-called robust band touchings \cite{bzdusek_robust_2017}. This centrosymmetric extension of the AZ classes list all the possible dimensionality of the nodes and the corresponding topological charges, which shows a interesting pattern of Bott periodicity as same as the AZ classes of gapped topological phases. The classifications can be found in a periodic table as represented in Table I of Ref. \onlinecite{bzdusek_robust_2017}.

One of our main goals here is to find a realistic material belonging to one of those gapless classes in the periodic table. By using first principle calculations, we can show that, at zero pressure \ch{YH_3} is a pseudo nodal surface topological semimetal belonging to the class BDI, for which the band crossing is protected by $\mathcal{T},\mathcal{I},\mathcal{C}$ and a approximate particle-hole $\mathcal{P}$ symmetry and the spin-orbital coupling is ignored due the fact that both Y and H are light elements. Without the particle-hole symmetry, the band crossings form a closed nodal ring belonging to class AI which may have a $\mathbf{Z}_2$ topological charge. The second goal of our work is to study the effect of the hydrostatic pressure on the electronic structure. We find that the increase of hydrostatic pressure doesn't affect the inversion symmetry but break the particle-hole symmetry gradually. Rather interestingly, the nodal ring keeps shrinking in the process and eventually gapped out, which indicates that the topological charge of the nodal ring is trivial.

We choose \ch{YH_3} as the object of study because yttrium hydrides are well known as they have many interesting properties under strain. The first principle calculation of structural and electronic properties of yttrium hydride has been obtained long time ago \cite{wang_structural_1995}. Later it was predicted that \ch{YH_3} undergoes a structural transformation by applying a pressure \cite{ahuja_semiconducting_1997}, and a lots of effort has been done to understand the structure of \ch{YH_3} \cite{palasyuk_pressure_2004,ohmura_infrared_2006,machida_x-ray_2006,remhof_hydrogen_1999,wolf_first-principles_2002}. Most recently, it was shown in Ref. \onlinecite{de_almeida_dynamical_2009} that \ch{YH_3} undergoes a pressure induced structural transformation at critical pressure $21$ GPa, which means that, at low pressure \ch{YH_3} has structure with space group $p\bar{3}c1$; above $21$ GPa, \ch{YH_3} is more stable with a cubic structure.

On the other hand, the electronic structure of \ch{YH_3} has also been reconsidered by means of first principle calculation recently. For instance, in Ref. \cite{shao_nonsymmorphic_2018} it was pointed out that hexagonal \ch{YH_3} is a topological semimetal with nodal line protected by the glide-plane symmetry. Later the authors in Ref. \onlinecite{wang_pseudo_2018} argue that \ch{YH_3} is actually a pseudo nodal surface semimetal protected by two mirror symmetry and inversion symmetry. By pseudo they mean that only the three nodal rings in the $k_{x(y,z)}=0$ planes are truly degenerated. Away from these nodal lines, the band gap is just approximately zero. Although these studies give different results, their method is quite similar. Most importantly, they all suggest that it is the crystalline symmetries that play a key role to protect the band crossings.

\begin{figure}[!h]
  \centering
    \begin{subfigure}[b]{0.45\textwidth}
      \includegraphics[width=\textwidth]{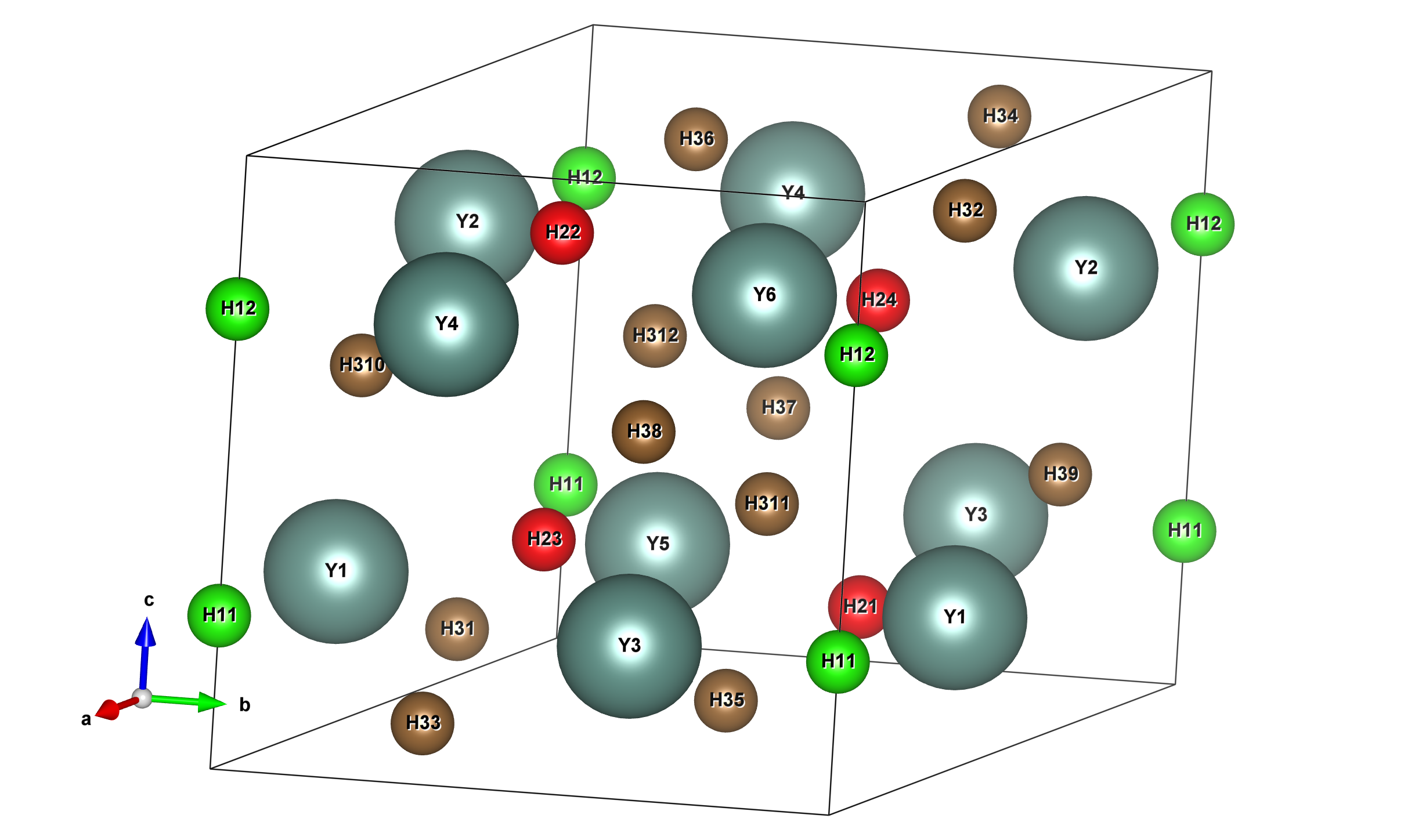}
      \caption{}
      \label{fig:crystall_structure}
      \end{subfigure}
  \begin{subfigure}[b]{0.45\textwidth}
      \includegraphics[width=\textwidth]{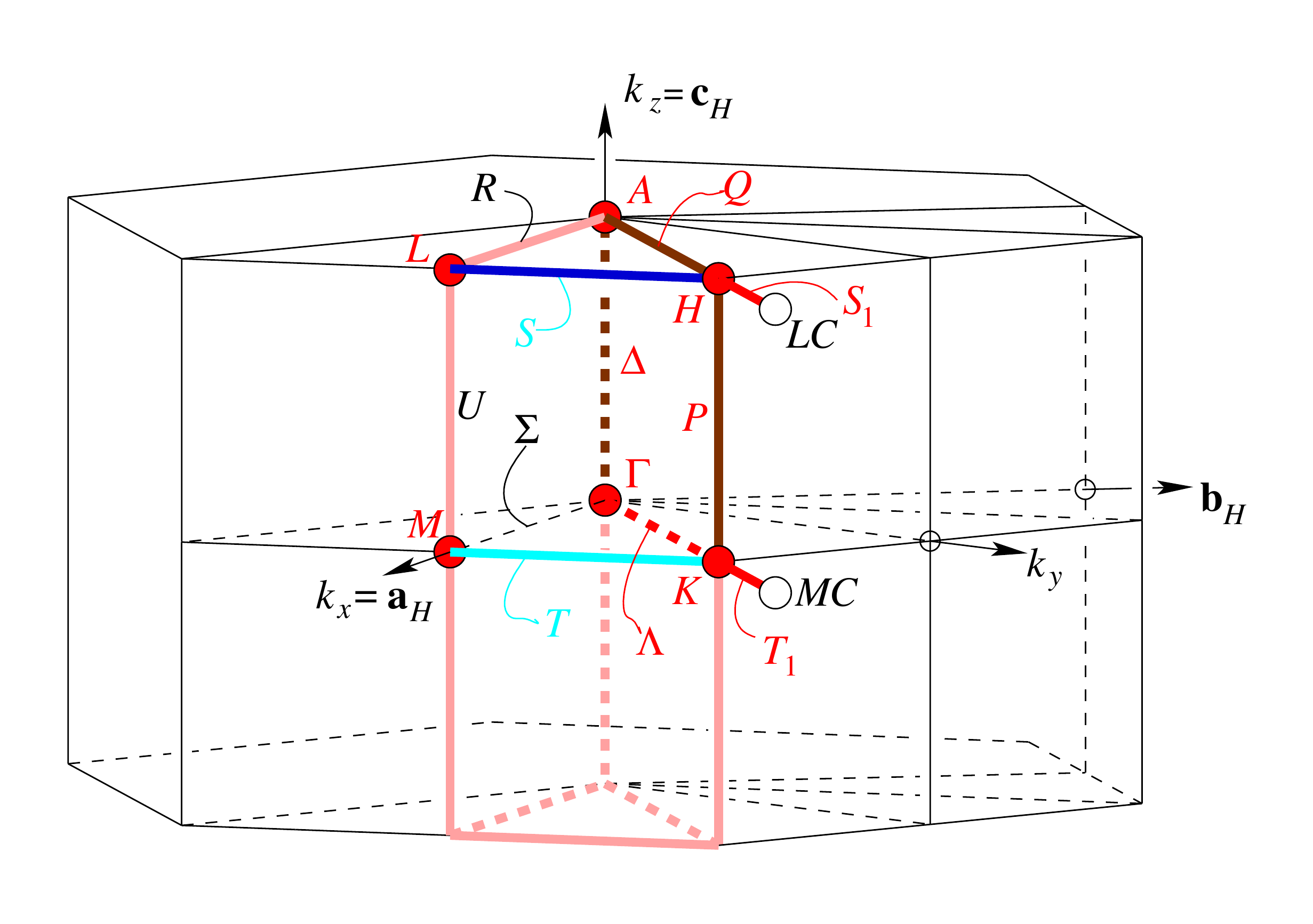}
      \caption{}
      \label{fig:Brillouin-zone}
  \end{subfigure}\\
  \begin{subfigure}[b]{0.55\textwidth}
    \includegraphics[width=\textwidth]{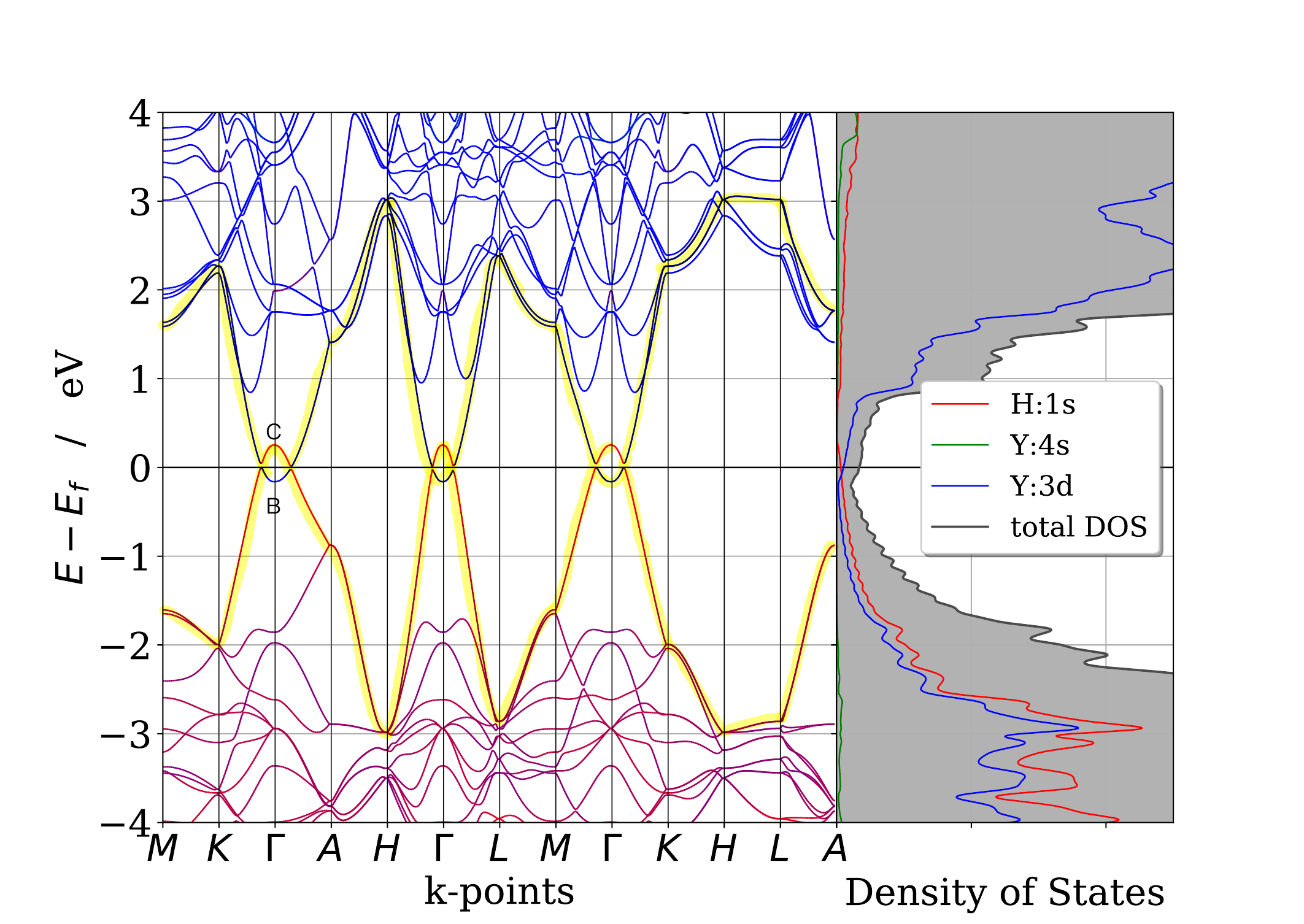}
    \caption{}
    \label{fig:banddos}
\end{subfigure}
\begin{subfigure}[b]{0.4\textwidth}
  \includegraphics[width=\textwidth]{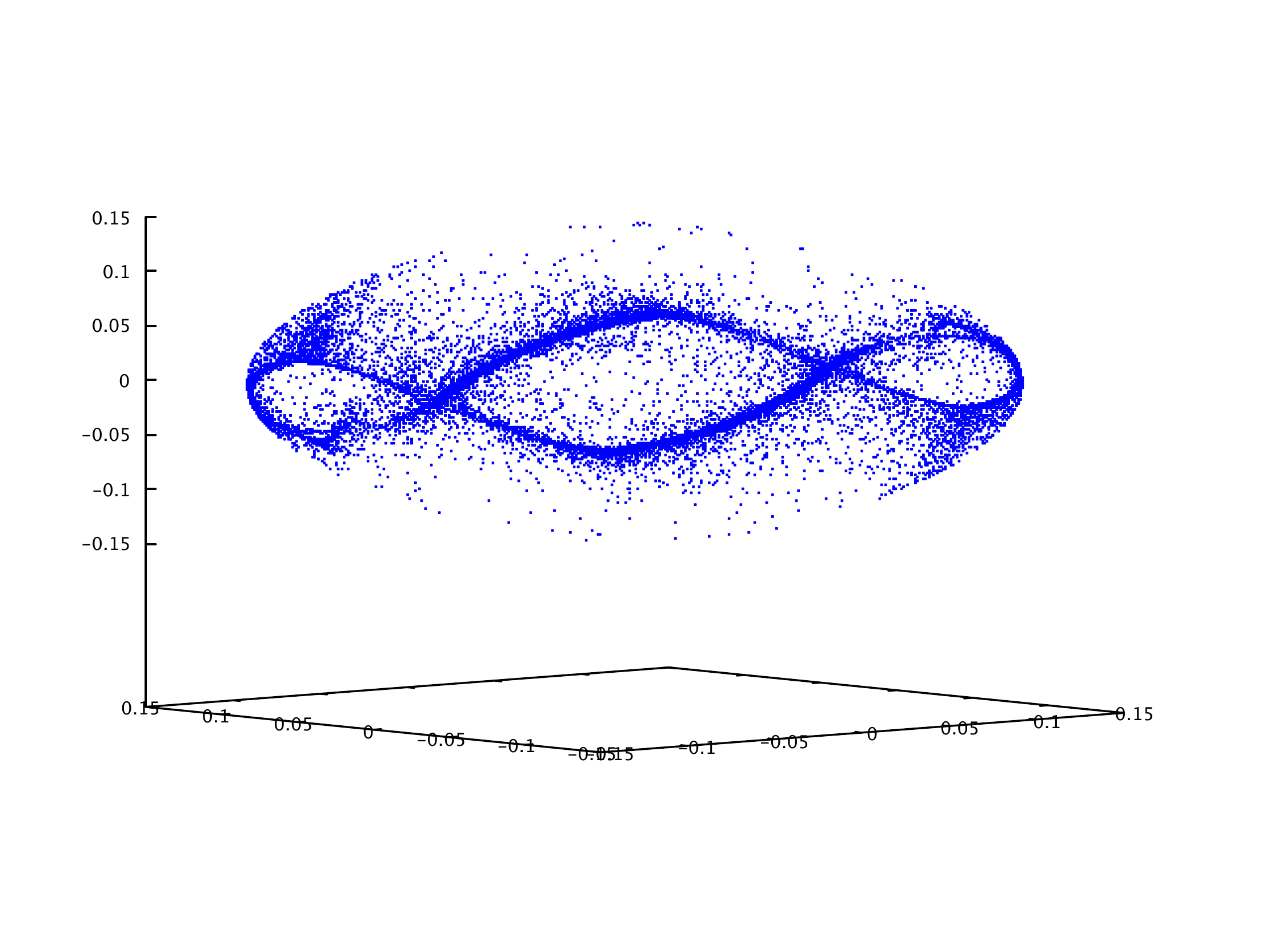}
  \caption{}
  \label{fig:nodes}
 \end{subfigure}
\caption{\raggedright The crystal structure and electronic structure of \ch{YH_3}. (a) The unit cell of \ch{YH_3} at zero pressure. There are $18$ H atoms belonging to $3$ different Wyckoff sets: $\{12g, 2a, 4d \}$, while all the $6$ Y atoms are equivalent. (b) The Brillouin zone (BZ). The high symmetry points and lines are labeled by following the convention of database \cite{i._aroyo_crystallography_2011,aroyo_bilbao_2006,aroyo_bilbao_2006-1}. (c) The band structure and density of state (DOS). (d) The scatter plot of the nodes around the $\Gamma$ point. The nodes can be found by comparing LUCB and the HOVB within a tiny range of error, i.e., searching for the k-points satisfying $|E_{LUCB}(k)-E_{HOVB}(k)|<0.005 $ eV. }
\end{figure}

The band structure of hexagonal \ch{YH_3} is revisited in this paper. However we suggest that the band crossings are protected by non-spatial symmetries rather than the local crystalline symmetries. The starting point is to choose a  crystal structure of \ch{YH_3}. According to Ref. \onlinecite{de_almeida_dynamical_2009}, \ch{YH_3} has the energy favored structure with symmetry group $p\bar{3}c1$ (No. 165) at zero pressure. The initial structure is obtained from the Materials Project \cite{persson_materials_2016-1}. As shown in FIG.~\ref{fig:crystall_structure}, in one unit cell there are three different sets of hydrogen atoms and six equivalent Y atoms. Under the symmetry operations listed in TABLE \ref{tab:generators}, these atoms transform to each other only within each equivalent set.

\begin{table}[h!]
  \begin{ruledtabular}
  \begin{tabular}{c c c c c c c}
    No. & $(x, y, z)$ form & Seitz symbol & \quad & No. & $(x, y, z)$ form & Seitz symbol \\ [0.5ex] 
    \hline
    $1$   &  $x,y,z$         & $\{1|0\}$                         & \quad & $7$         &    $-x,-y,-z $   & $\{ -1 | 0 \}$  \\
    $2$   &  $-y,x-y,z$      & $\{ 3^{+}_{001} | 0 \} $          & \quad &$8$       &  $   y,-x+y,-z    $   & $\{ -3^{+}_{001} | 0 \} $  \\
    $3$   & $-x+y,-x,z  $    & $ \{ 3^{-}_{001} | 0 \}  $        &  \quad & $9$     & $x-y,x,-z $    & $ \{ -3^{-}_{001} | 0 \}$\\
    $4$   &  $-y,-x,z+1/2$  & $\{ m_{110} | 0 0 \frac{1}{2} \}$  &  \quad & $10$       &  $y,x,-z+1/2$     & $\{ 2_{110} | 0 0 \frac{1}{2} \}$ \\
    $5$   &  $-x+y,y,z+1/2$  & $\{ m_{100} | 0 0 \frac{1}{2} \}$ &  \quad &$11$       & $x-y,-y,-z+1/2$  & $\{ 2_{100} | 0 0 \frac{1}{2} \}$\\ 
    $6$   &  $ x,x-y,z+1/2$  & $\{ m_{010} | 0 0 \frac{1}{2} \}$ &  \quad &$12$      & $ -x,-x+y,-z+1/2$  & $\{ 2_{010} | 0 0 \frac{1}{2} \}$
    \end{tabular}
    \caption{The generators of space group $p\bar{3}c1$.}
\label{tab:generators}
  \end{ruledtabular}
  \end{table}

Given the structure parameters of \ch{YH_3}, we perform the first principle calculation based on density-functional theory \cite{hohenberg_inhomogeneous_1964,kohn_self-consistent_1965} implemented by the Vienna ab initio simulation package (VASP) \cite{kresse_efficient_1996,kresse_efficiency_1996}. The projector augmented wave (PAW) method \cite{blochl_projector_1994,kresse_ultrasoft_1999} is used with the Perdew-Burke-Ernzerhof (PBE) realization, where the generalized gradient approximation (GGA) is adopted for the exchange-correlation potential \cite{perdew_generalized_1996}. The reciprocal space sampling was performed using $25 \times 25 \times 25$ Monkhorst-Pack meshes \cite{monkhorst_special_1976}. Optimization of structural parameters was achieved by a minimization of atomic forces and stress tensors applying the conjugate gradient technique. 

As shown in FIG. \ref{fig:banddos}, the accidental band crossings have emerged at the fermi level because of the overlapping of an electron- and hole-like band, we call them e-band and h-band accordingly. At $\Gamma$ point, they happen to be the highest occupied valence band (HOVB) and the lowest unoccupied conducting band (LUCB). Indeed, when closer to the $\Gamma$ point, they are the only two bands near the fermi level which are well separated from other bands. Further away from the $\Gamma$ point, the e-band and h-band will cross over other bands. Especially on the edges of the BZ, they will degenerate with other bands due to the crystalline symmetries. To understand the global pattern of the band structure is not a trivial task, in general,  one may need to use more powerful combinatorial methods involving $K$-theory \cite{kruthoff_topological_2017} or graph theory \cite{bradlyn_topological_2017,vergniory_graph_2017,bradlyn_band_2018}. For our case, we only care about the e-band and h-band, this can be done by following the rules of compatibility relations, which is the branching rules of the irreducible representations (irreps) of the little group at different high symmetric points. This is to say, we need to analyse the band connectivity for the two bands only. It can be done by firstly perform the first principle calculation to obtain the wave functions, and then use the BANDREP program on the Bilbao Crystallographic Server \cite{i._aroyo_crystallography_2011,aroyo_bilbao_2006,aroyo_bilbao_2006-1} to analyse how the wave functions change under the crystalline symmetry transformation. In this way, we obtain the irreps for the bands near the fermi level and compile them in TABLE.~\ref{tab:rep}. Also, the branching rules of these irreps are listed in TABlE.~\ref{tab:branching}.

Since the e-band and h-band are not degenerated at the $\Gamma$ point, if one starts from the irreps at $\Gamma$ point, following the branching rules, go along a close k-path back to the $\Gamma$ point, one should arrive the same irreps as it was started. For example, let's find out the band connectivity for h-band along the k-path $\Gamma-K-H-L-A-\Gamma$: From the branching rules list in TABlE.~\ref{tab:branching}, it is not hard to verify that the irreps $\Gamma_2^--K_3-H_3-L_1-A_3-\Gamma_2^-$ along the k-path is compatible with the branching rules. 
On the other hand, the bands are already labelled by the irreps as shown in TABLE.~\ref{tab:rep}. This information is enough for us to figure out the connectivity of h-band. We apply the same method to e-band and highlight the connectivity pattern with yellow color as show in FIG. \ref{fig:banddos}.
\begin{table}[!h]
  \begin{ruledtabular}
  \begin{tabular}{ c c c c c c c }
  
  \textbf{Bands} & \textbf{M} & \textbf{$\Gamma$}                & \textbf{A}                   & \textbf{H}                   & \textbf{K}                         & \textbf{L}                         \\ \hline
  LUCB+5         & $M_2^+$    & \multirow{2}{*}{$\Gamma_3^+$}    & \multirow{4}{*}{$A_1A_2(4)$} & \multirow{4}{*}{$H_3H_3(4)$} & $K_1$                     & \multirow{2}{*}{$L_1(2)$} \\ 
  LUCB+4         & $M_1^-$    &                                  &                              &                              & \multirow{2}{*}{$K_3(2)$} &                           \\ 
  LUCB+3         & $M_1^+$    & $\Gamma_1^-$                     &                              &                              &                           & \multirow{2}{*}{$L_1(2)$} \\
  LUCB+2         & $M_2^-$    & \multirow{2}{*}{$\Gamma_3^-(2)$} &                              &                              & \multirow{2}{*}{$K_3(2)$} &                           \\  
  LUCB+1         & $M_1^-$    &                                & \multirow{2}{*}{$A_3(2)$}    & \multirow{2}{*}{$H_1H_2(2)$} &                           & \multirow{2}{*}{$L_1(2)$} \\ 
  LUCB           & $M_2^+$    & $\Gamma_2^-$                     &                              &                              & $K_1$                     &                           \\ 
    \hline
  HOVB           & $M_2^-$    & $\Gamma_2^+$                     & \multirow{2}{*}{$A_3(2)$}    & \multirow{4}{*}{$H_3H_3(4)$} & \multirow{2}{*}{$K_3(2)$} & \multirow{2}{*}{$L_1(2)$} \\ 
  HOVB-1         & $M_1^+$    & $\Gamma_1^-$                     &                              &                              &                           &                           \\
  HOVB-2         & $M_2^+$    & $\Gamma_1^+$                     & \multirow{2}{*}{$A_3(2)$}    &                              & $K_2$                     & \multirow{2}{*}{$L_1(2)$} \\ 
  HOVB-3         & $M_1^-$    & $\Gamma_2^+$                     &                              &                              & $K_3(2)$                  &                           \\ 
  \end{tabular}
\end{ruledtabular}
  \caption{\raggedright Band representations near the fermi surface. We adopt the conventions of the Bilbao Crystallographic Server \cite{i._aroyo_crystallography_2011,aroyo_bilbao_2006,aroyo_bilbao_2006-1} to name the irreps. The number in the parentheses indicates the degree of degeneracy.}
  \label{tab:rep}
  \end{table}

  \begin{table}[!h]
    \begin{ruledtabular}
    \begin{tabular}{c l cl c l}
    
    \multicolumn{2}{c}{\textbf{$\mathbf{\Gamma}$--$\mathbf{\Lambda}$--K}}                    & \multicolumn{2}{c}{\textbf{H--P--K}}                & \multicolumn{2}{c}{\textbf{K--T--M}}                              \\ \hline
    \multicolumn{2}{c}{$\Gamma_1^\pm \rightarrow \Lambda_1$}              & \multicolumn{2}{c}{$H_1\rightarrow P_1$}            & \multicolumn{2}{c}{$K_1\rightarrow T_1$}                          \\
    \multicolumn{2}{c}{$\Gamma_2^\pm \rightarrow \Lambda_2$}              & \multicolumn{2}{c}{$H_2\rightarrow P_1$}            & \multicolumn{2}{c}{$K_2\rightarrow T_2$}                          \\
    \multicolumn{2}{c}{$\Gamma_3^\pm(2) \rightarrow \Lambda_1+\Lambda_2$} & \multicolumn{2}{c}{$H_3(2)\rightarrow P_2P_3(2)$}        & \multicolumn{2}{c}{$K_3(2)\rightarrow T_1+T_2$}                   \\
    \multicolumn{2}{c}{$K_1 \rightarrow \Lambda_1$}                       & \multicolumn{2}{c}{$K_1\rightarrow P_1$}            & \multicolumn{2}{c}{$M_2^\pm \rightarrow T_2$}                      \\
    \multicolumn{2}{c}{$K_2 \rightarrow \Lambda_2$}                       & \multicolumn{2}{c}{$K_2\rightarrow P_1$}            & \multicolumn{2}{c}{}                     \\
    \multicolumn{2}{c}{$K_3(2) \rightarrow \Lambda_1+\Lambda_2$}          & \multicolumn{2}{c}{$K_3(2)\rightarrow P_2P_3(2)$}        & \multicolumn{2}{c}{}                                              \\ \hline
    \multicolumn{2}{c}{\textbf{H--Q--A}}                                  & \multicolumn{2}{c}{\textbf{H--S--L}}                & \multicolumn{2}{c}{\textbf{L--R--A}}                              \\ \hline
    \multicolumn{2}{c}{$H_{1}H_2(2) \rightarrow Q_1Q_2(2)$}                   & \multicolumn{2}{c}{$H_{1}H_2(2)  \rightarrow S_1S_2(2)$} & \multicolumn{2}{c}{$L_1(2)\rightarrow R_1+R_2$}                   \\
    \multicolumn{2}{c}{$H_3(2) \rightarrow Q_1Q_2(2) $}                   & \multicolumn{2}{c}{$H_3(2) \rightarrow S_1S_2(2) $} & \multicolumn{2}{c}{$A_{1(2,3)}(2)\rightarrow R_1+R_2$}               \\
    \multicolumn{2}{c}{$A_3(2)\rightarrow Q_1Q_2(2)$}                     & \multicolumn{2}{c}{$L_1(2)\rightarrow S_1S_2(2)$}    & \multicolumn{2}{c}{}                                              \\ \hline
    \multicolumn{2}{c}{\textbf{$\mathbf{\Gamma}$--$\mathbf{\Sigma}$--M}}                    & \multicolumn{2}{c}{\textbf{L--U--M}}                & \multicolumn{2}{c}{\textbf{$\mathbf{\Gamma}$--$\mathbf{\Delta}$--A}}                \\ \hline
    \multicolumn{2}{c}{$\Gamma_1^+, \Gamma_2^- \rightarrow \Sigma_1$}     & \multicolumn{2}{c}{$L_1(2)\rightarrow U_1+U_2$}     & \multicolumn{2}{c}{$\Gamma_1^+, \Gamma_2^- \rightarrow \Delta_1$} \\
    \multicolumn{2}{c}{$\Gamma_1^-, \Gamma_2^+ \rightarrow \Sigma_2$}     & \multicolumn{2}{c}{$M_1^+, M_2^- \rightarrow U_1$}  & \multicolumn{2}{c}{$\Gamma_1^-, \Gamma_2^+ \rightarrow \Delta_2$} \\
    \multicolumn{2}{c}{$\Gamma_3^\pm (2)\rightarrow \Sigma_1+\Sigma_2$}   & \multicolumn{2}{c}{$M_1^-, M_2^+ \rightarrow U_2$}  & \multicolumn{2}{c}{$\Gamma_3^\pm (2)\rightarrow \Delta_3(2)$}     \\
    \multicolumn{2}{c}{$M_1^+, M_2^- \rightarrow \Sigma_1$}               & \multicolumn{2}{c}{}                                & \multicolumn{2}{c}{$A_1(2)\rightarrow \Delta_3(2)$}               \\
    \multicolumn{2}{c}{$M_1^-, M_2^+ \rightarrow \Sigma_2$}               & \multicolumn{2}{c}{}                                & \multicolumn{2}{c}{$A_2(2)\rightarrow \Delta_3(2)$}               \\
    \multicolumn{2}{c}{}                                                  & \multicolumn{2}{c}{}                                & \multicolumn{2}{c}{$A_3(2)\rightarrow \Delta_1+\Delta_2$}         \\ 
    
    \end{tabular}
  \end{ruledtabular}
    \caption{All possible branching rules of irreps between different high symmetry points of the little groups of $p\bar{3}c1$.}
    \label{tab:branching}
  \end{table}

Before we say anything about the symmetries of the two bands system, first notice that the band crossings are located away from the high-symmetry points,so if one wants to use the Dirac-matrix Hamiltonian method to analyze the band structure, one has to take into account the full momentum dependence of the Hamiltonian in the entire BZ. Therefore the symmorphic crystalline symmetries such as rotations and reflections located at the $\Gamma$ point will not be enough to constrain the Hamiltonian form. This is a hint that one should be looking for the global symmetries, such as $\mathcal{T},\mathcal{P}$ and $\mathcal{C}$. These non-spatial symmetries requires that the Hamiltonian $\mathcal{H}(k)$ in momentum space must satisfy
\begin{equation}
  \label{eq:non-spatial}
  \begin{split}
  \mathcal{T}\mathcal{H}(k)\mathcal{T}^{-1}&=\mathcal{H}(-k),  \quad \mathcal{T}^2 =\pm 1\\
  \mathcal{P}\mathcal{H}(k)\mathcal{P}^{-1}&=-\mathcal{H}(-k),  \quad \mathcal{P}^2 =\pm 1\\
  \mathcal{C}\mathcal{H}(k)\mathcal{C}^{-1}&=-\mathcal{H}(k),  \quad \mathcal{C}^2 = 1
  \end{split}
 \end{equation}
Note that $\mathcal{T}$ and $\mathcal{P}$ will yield $\mathcal{C}$
\begin{equation}
  \mathcal{C}=\mathcal{T}\cdot \mathcal{P}.
\end{equation}
However the inverse is not true, i.e., $\mathcal{C}$ doesn't ensure the existence of $\mathcal{T}$ and $\mathcal{P}$. It is the ten possible combinations of these symmetries that lead to the ten-fold way classification of gapped Hamiltonian $\mathcal{H}(k)$ \cite{schnyder_classification_2008,kitaev_periodic_2009,ryu_topological_2010}. For our case, which is a gapless systems, as suggested by Ref. \onlinecite{bzdusek_robust_2017}, one can achieve a similar classification by taking into account the inversion symmetry $\mathcal{I}$. In a nutshell, one can introduce the new symmetry operators
\begin{equation}
  \mathbb{P} \equiv \mathcal{P}\mathcal{I}, \quad \mathbb{T}= \mathcal{TI}, 
\end{equation}
which act on Hamiltonian as
\begin{equation}
  \begin{split}
    \mathbb{T}\mathcal{H}(k)\mathbb{T}^{-1}&=\mathcal{H}(k), \quad  \mathbb{T}^2 =\pm 1,\\
    \mathbb{P}\mathcal{H}(k)\mathbb{P}^{-1}&=-\mathcal{H}(k),  \quad \mathbb{P}^2 =\pm 1.
  \end{split}
 \end{equation}
 They are also anti-unitary and square to $\pm 1$ similar to $\mathcal{P}$ and $\mathcal{T}$. But now they don't act on the momentum space globally, that is to say, relate Hamiltonians at different $k$-points such as $-k$ and $k$. They can be seen as the localized version of $\mathcal{P}$ and $\mathcal{T}$, by that we mean they are now the symmetries of the Hamiltonian on the entire BZ rather than only on the time reversal invariant momenta (TRIM). The classification of gapless Hamiltonians based on $\mathbb{T},\mathbb{P}$ and $\mathcal{C}$ can be found in Ref. \cite{bzdusek_robust_2017}, in which the dimensionality and topological invariants of the nodes for each AZ$+\mathcal{I}$ classes are also given explicitly.

Knowing that \ch{YH_3} already has the time reversal symmetry and inversion symmetry, for now the major problem is to determine whether or not $\mathcal{P}$ exists, if so, then the chiral symmetry is automatically guaranteed. For a system with particle-hole symmetry, the band structure will have a recognizable feature: the spectrum must be symmetric around the Fermi level. Indeed, for every state $\psi$ with energy $E$, there will be a particle-hole symmetric state $\mathcal{P}\psi $ with energy $-E$. As one can see, the two bands are approximately mirrored by the fermi level. This is not an accident, but rather a signature of approximate particle-hole symmetry. One can also understand this phenomenon from the view of quantum chemistry. In FIG. \ref{fig:banddos}, the projected density of states shows that, the valence bands are mostly contributed by H-s orbital and the occupied Y-d orbital. These two orbitals are strongly coupled, which is exactly the configuration for the valence electrons of \ch{YH_3}. On the other hand, the conduction bands are mostly contributed by the unoccupied Y-d orbitals. To put it another way, H atom contributes one s electron and Y contributes three d electrons to form the bond of \ch{YH_3}. It is mostly the two unoccupied Y-d orbitals which generate the conduction bands. The bonded electrons of H atoms can be viewed as a Dirac sea. When they acquire enough energy, it is possible for them to jump onto the unoccupied Y-d orbitals. This will creates a hole in the Dirac sea. The overlap of the two bands and the particle-hole symmetry between them are exactly the consequence of the jumping.

It is clear now, if we take the imperfection of the particle-hole symmetry as a small effect, the band crossings can be considered as class BDI protected by $\mathbb{T},\mathbb{P}$ and $\mathcal{C}$ symmetries. Here we should emphasize that the particle-hole symmetry is only an approximate symmetry, the nodes are not a perfectly surface but with a small gap opened due to the weakly broken $\mathbb{P}$ symmetry. On the other hand, the time reversal symmetry is always exact, which makes sure that the band crossings always contain a nodal line and fall within class AI. We borrow the word "pseudo" from Ref. \onlinecite{wang_pseudo_2018} and call the band crossings of \ch{YH_3} a pseudo nodal surface of BDI class. Actually we can plot the nodes directly from first principle calculation: By using maximally localized Wannier functions implemented by the Wannier90 package \cite{mostofi_updated_2014} and the WannierTools package \cite{wu_wanniertools_2018}, we can find all the nodes by comparing the LUCB and the HOVB within a tiny range of error, that is, the nodes can be defined as all the k-points satisfying $E_{LUCB}(\mathbf{k})-E_{HOVB}(\mathbf{k})<E_{error}$. This error indicates how many gap opened due to the broken $\mathbb{P}$ symmetry one can tolerate. As shown in FIG. \ref{fig:nodes}, we have chosen the error to be $E_{error}=0.005$ eV, the result shows that there is a closed nodal line on a fuzzy surface surrounding the $\Gamma$ point at the center. Away from the line, the k-points meet the demand of $E_{LUCB}(\mathbf{k})-E_{HOVB}(\mathbf{k})<E_{error}$ become sparse. As we turn down the value of $E_{error}$, the number of nodes away from the line become fewer and fewer. For now, we think $0.005$ eV is smaller enough and the pseudo nodal surface is quite obvious from FIG.~\ref{fig:nodes}. In conclusion, we could say that \ch{YH_3} is a very special nodal line semimetal belonging to class AI with a slightly broken particle-hole symmetry, within certain error tolerance, \ch{YH_3} can also be viewed as a nodal surface semimetal belonging to class BDI.

Indeed, without spin-orbit coupling, one can obtain the effective two bands Hamiltonian. Following Ref.~\cite{bzdusek_robust_2017}, the symmetry operators $\mathbb{T},\mathbb{P}$ and $\mathcal{C}$ can be represented as
\begin{equation}
  \label{eq:rep-operators}
  \mathbb{T}=\mathcal{K}, \quad \mathbb{P}=\sigma_z\mathcal{K},\quad \mathcal{C}=\sigma_z,
\end{equation}
where $\sigma_z$ is the Dirac matrix and $\mathcal{K}$ indicates complex conjugation. The Hamiltonian can be given by
\begin{equation}
  \label{eq:halmiltonian}
  \mathcal{H}(\mathbf{k})=f(\mathbf{k})\sigma_z +g(\mathbf{k})\sigma_x.
\end{equation}
Remember that $\mathbb{P}$ is an approximately symmetry, only $\mathbb{T}$ is a rigid symmetry, therefore the $f(\mathbf{k})$ term in \eqref{eq:halmiltonian} which breaks symmetry $\mathbb{P}$ can be seen as a small perturbation. Turn off $f(\mathbf{k})$, the resulting nodes are just a surface represented by $g(\mathbf{k})=0$, which exactly belongs to the class BDI. With $f(\mathbf{k})\neq 0$, the resulting nodes are a closed nodal line on surface $g(\mathbf{k})=0$ satisfying $f(\mathbf{k})=0$, thus belongs to the class AI. 

We have more to say about the particle-hole symmetry of \ch{YH_3}, which is, $\mathbb{P}$ is pressure sensitive. It is not a surprise, though, that hydrogen-rich materials are well known for having attractive properties under pressure, for example, they could exhibit high-temperature superconductivity under high pressure \cite{ashcroft_hydrogen_2004,liu_high-pressure_2017,liu_potential_2017,peng_hydrogen_2017}. Making effort to fully understand this phenomenon is still ongoing. Here we just consider a very specific scenario that \ch{YH_3} is subjected to hydrostatic pressure only. The reason is that hydrostatic pressure will not change the space inversion symmetry, but can cause the lattice constants to be decreased and induces changes in the electronic structure \cite{keyes_effects_1960}. 

\begin{figure}[!h]
  \centering
    \begin{subfigure}[b]{0.45\textwidth}
      \includegraphics[width=\textwidth]{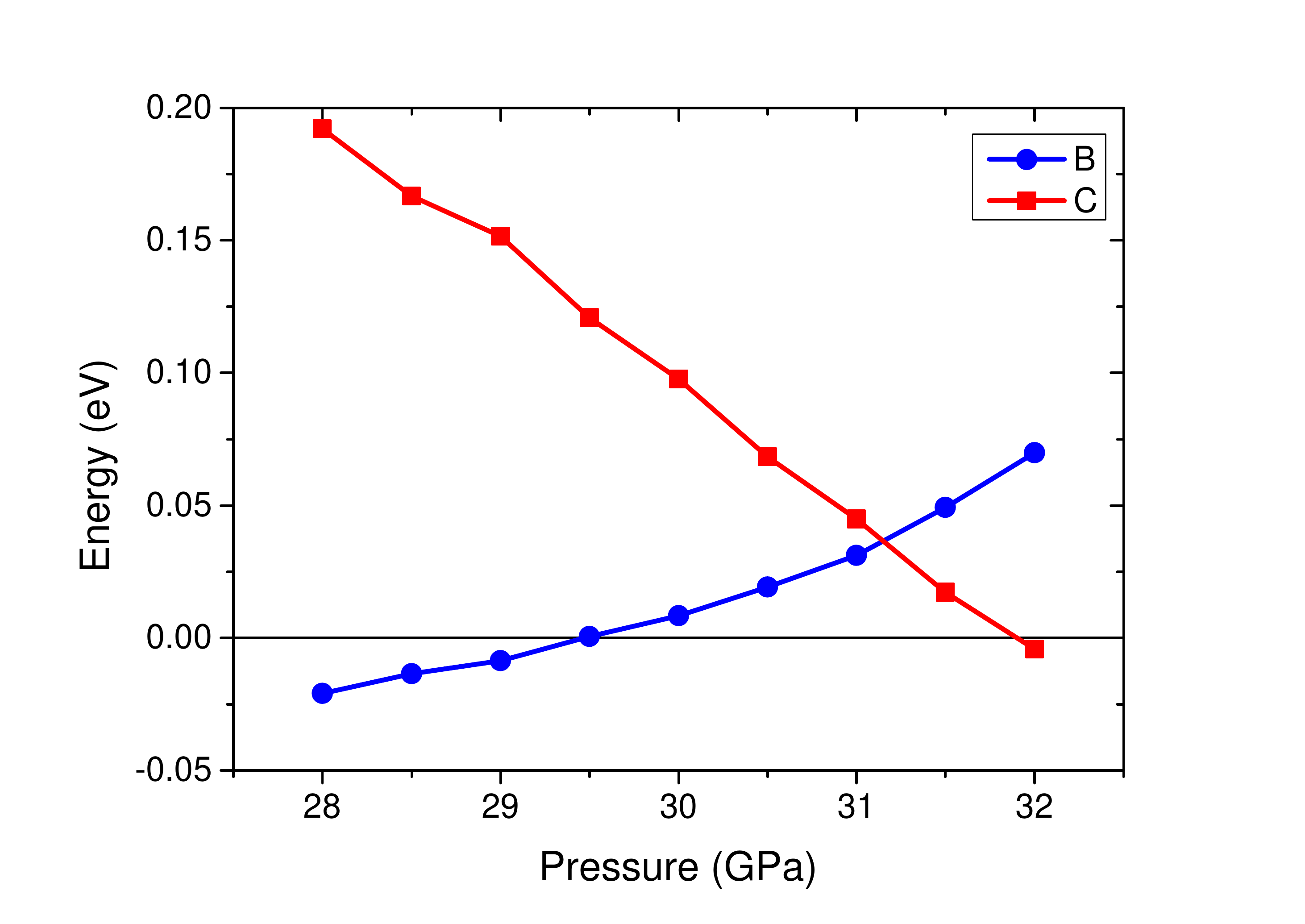}
      \caption{}
      \label{fig:pressure}
      \end{subfigure}
  \begin{subfigure}[b]{0.45\textwidth}
      \includegraphics[width=\textwidth]{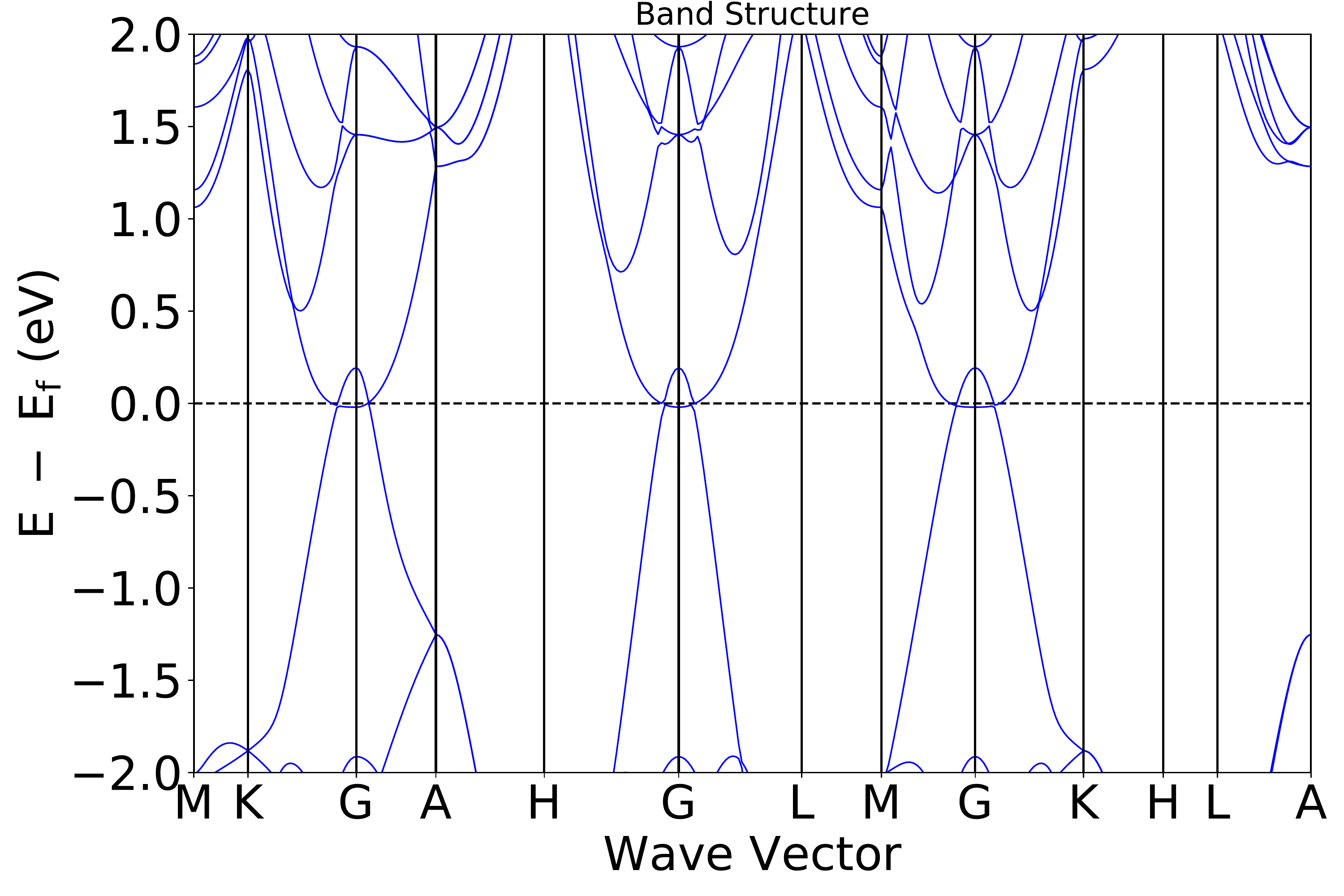}
      \caption{}
      \label{fig:280}
  \end{subfigure}\\
  \begin{subfigure}[b]{0.45\textwidth}
    \includegraphics[width=\textwidth]{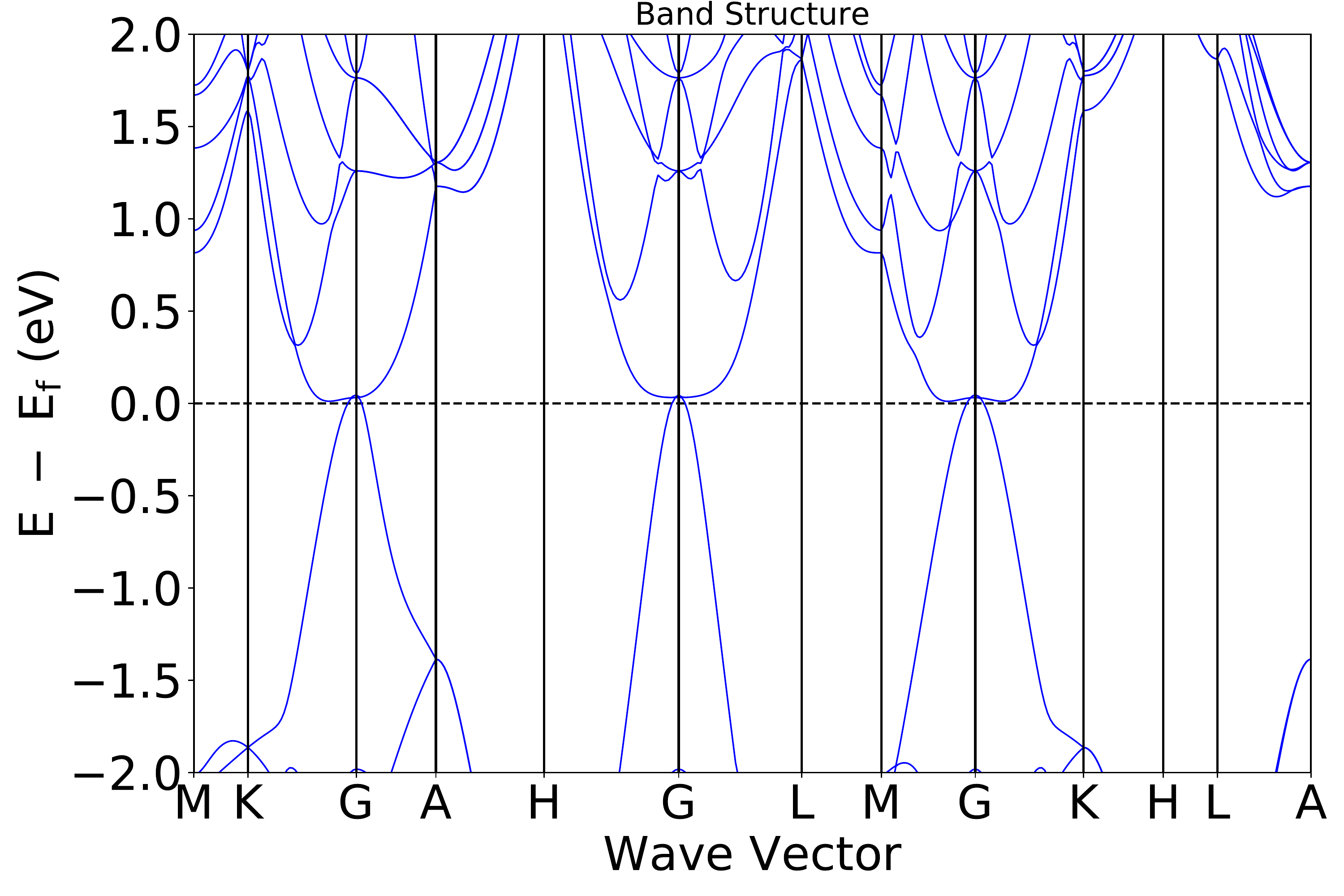}
    \caption{}
    \label{fig:310}
    \end{subfigure}
\begin{subfigure}[b]{0.45\textwidth}
    \includegraphics[width=\textwidth]{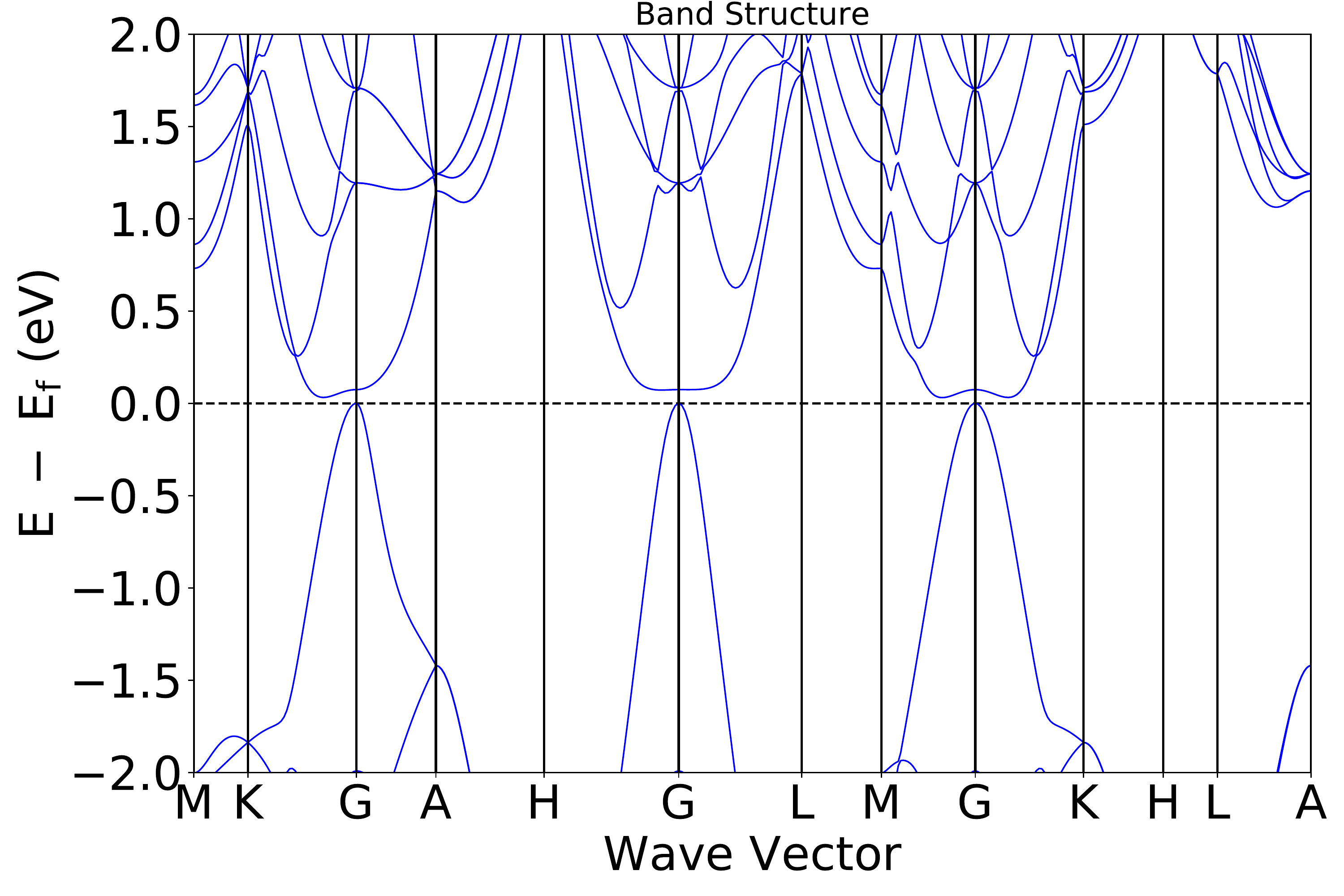}
    \caption{}
    \label{fig:320}
\end{subfigure}
\caption{\raggedright Pressure induced electronic structure changes. (a) The highest energy of h-band and lowest energy of e-band change with pressure increasing. (b) Band structure at $28$ GPa. (c) Band structure at $31$ GPa. (d) Band structure at $32$ GPa.}
\end{figure}

Our calculation shows that the particle-hole symmetry appears at zero pressure will be broken when the pressure is increased. The fundamental reason is that the highest energy of h-band and the lowest energy of e-band, denoted by $C$ and $B$ in FIG.~\ref{fig:banddos} respectively, vary differently when increasing the pressure. As shown in FIG.~\ref{fig:pressure} and FIG.~\ref{fig:280}, at $28$ GPa, the energy at $B$ and $C$ points are not mirrored across the fermi level, which indicates that the particle-hole symmetry is broken. Thus many points on the nodal surface will be gapped out expect a closed nodal line protected by the time reversal symmetry and inversion symmetry. This kind of nodal line of class AI has a $\mathbf{Z}_2$ topological charge, which is defined by the Berry’s phase for all occupied bands: either $0$ or $\pi$ \cite{fang_topological_2015,bzdusek_robust_2017}, corresponding to trivial and the nontrivial cases accordingly.

To figure out whether the nodal ring possess a non-trivial topological charge or not, one can smoothly increase the pressure and observe how the band structure changes. As one can see, from $28$ GPa to $31$ GPa, as shown in FIG.~\ref{fig:280} and FIG.~\ref{fig:310}, the nodal ring keeps shrinking as the pressure increasing. Up to about $31$ GPa, as shown in FIG. \ref{fig:310}, the nodal ring shrinks to a point, notice that this actually agreed with the crossing point in FIG. \ref{fig:pressure}, where the highest energy of h-band and the lowest energy of e-band coincide. When the pressure goes up to $32$ GPa as shown in FIG. \ref{fig:320}, there is no overlap of e-band and h-band, the nodal ring is fully gapped out without encountering any sudden changes. This phenomenon indicates that the nodal ring is topological trivial, i.e., the Berry's phase for all occupied bands is quantized to $0$ not $\pi$ \cite{fang_topological_2015}.

We have studied the electronic structure of hexagonal \ch{YH_3} without spin-orbital coupling. At zero pressure, the system has space inversion symmetry, time reversal symmetry and an approximate particle-hole symmetry. The band crossings can be viewed as a pseudo nodal surface belonging to class BDI of zhe AZ+$\mathcal{I}$ classification \cite{bzdusek_robust_2017}. By pseudo we mean there actually exist a small gap about $0.005$ eV away from the nodal ring which belongs to class AI. We also find that the approximate particle-hole symmetry is pressure dependent. As long as the hydrostatic pressure is increasing, until it reaches $31$ GPa, the particle-hole symmetry will be gradually broken and the small gap on the pseudo nodal surface will be enlarged except for the residue nodal ring. In the meantime, the nodal ring keeps shrinking. When the pressure exceeds $31$ GPa, the nodal ring shrinks to a points and finally gapped out, which indicates that the nodal ring possesses a trivial $\mathbf{Z}_2$ topological invariant. This could be viewed as a topological phase transformation induced by pressure changes.

We thank Prof. Shaoyi Wu of University of University of Electronic Science and Technology of China for his help in VASP calculations.

\bibliography{YH3}

\begin{thebibliography}{46}%
\makeatletter
\providecommand \@ifxundefined [1]{%
 \@ifx{#1\undefined}
}%
\providecommand \@ifnum [1]{%
 \ifnum #1\expandafter \@firstoftwo
 \else \expandafter \@secondoftwo
 \fi
}%
\providecommand \@ifx [1]{%
 \ifx #1\expandafter \@firstoftwo
 \else \expandafter \@secondoftwo
 \fi
}%
\providecommand \natexlab [1]{#1}%
\providecommand \enquote  [1]{``#1''}%
\providecommand \bibnamefont  [1]{#1}%
\providecommand \bibfnamefont [1]{#1}%
\providecommand \citenamefont [1]{#1}%
\providecommand \href@noop [0]{\@secondoftwo}%
\providecommand \href [0]{\begingroup \@sanitize@url \@href}%
\providecommand \@href[1]{\@@startlink{#1}\@@href}%
\providecommand \@@href[1]{\endgroup#1\@@endlink}%
\providecommand \@sanitize@url [0]{\catcode `\\12\catcode `\$12\catcode
  `\&12\catcode `\#12\catcode `\^12\catcode `\_12\catcode `\%12\relax}%
\providecommand \@@startlink[1]{}%
\providecommand \@@endlink[0]{}%
\providecommand \url  [0]{\begingroup\@sanitize@url \@url }%
\providecommand \@url [1]{\endgroup\@href {#1}{\urlprefix }}%
\providecommand \urlprefix  [0]{URL }%
\providecommand \Eprint [0]{\href }%
\providecommand \doibase [0]{http://dx.doi.org/}%
\providecommand \selectlanguage [0]{\@gobble}%
\providecommand \bibinfo  [0]{\@secondoftwo}%
\providecommand \bibfield  [0]{\@secondoftwo}%
\providecommand \translation [1]{[#1]}%
\providecommand \BibitemOpen [0]{}%
\providecommand \bibitemStop [0]{}%
\providecommand \bibitemNoStop [0]{.\EOS\space}%
\providecommand \EOS [0]{\spacefactor3000\relax}%
\providecommand \BibitemShut  [1]{\csname bibitem#1\endcsname}%
\let\auto@bib@innerbib\@empty
\bibitem [{\citenamefont {Bzdusek}\ and\ \citenamefont
  {Sigrist}(2017)}]{bzdusek_robust_2017}%
  \BibitemOpen
  \bibfield  {author} {\bibinfo {author} {\bibfnamefont {T.}~\bibnamefont
  {Bzdusek}}\ and\ \bibinfo {author} {\bibfnamefont {M.}~\bibnamefont
  {Sigrist}},\ }\href {\doibase 10.1103/PhysRevB.96.155105} {\bibfield
  {journal} {\bibinfo  {journal} {Physical Review B}\ }\textbf {\bibinfo
  {volume} {96}} (\bibinfo {year} {2017}),\
  10.1103/PhysRevB.96.155105}\BibitemShut {NoStop}%
\bibitem [{\citenamefont {Fu}\ \emph {et~al.}(2007)\citenamefont {Fu},
  \citenamefont {Kane},\ and\ \citenamefont {Mele}}]{fu_topological_2007}%
  \BibitemOpen
  \bibfield  {author} {\bibinfo {author} {\bibfnamefont {L.}~\bibnamefont
  {Fu}}, \bibinfo {author} {\bibfnamefont {C.~L.}\ \bibnamefont {Kane}}, \ and\
  \bibinfo {author} {\bibfnamefont {E.~J.}\ \bibnamefont {Mele}},\ }\href
  {\doibase 10.1103/PhysRevLett.98.106803} {\bibfield  {journal} {\bibinfo
  {journal} {Phys. Rev. Lett.}\ }\textbf {\bibinfo {volume} {98}},\ \bibinfo
  {pages} {106803} (\bibinfo {year} {2007})}\BibitemShut {NoStop}%
\bibitem [{\citenamefont {Hasan}\ and\ \citenamefont
  {Kane}(2010)}]{hasan_colloquium_2010}%
  \BibitemOpen
  \bibfield  {author} {\bibinfo {author} {\bibfnamefont {M.~Z.}\ \bibnamefont
  {Hasan}}\ and\ \bibinfo {author} {\bibfnamefont {C.~L.}\ \bibnamefont
  {Kane}},\ }\href {\doibase 10.1103/RevModPhys.82.3045} {\bibfield  {journal}
  {\bibinfo  {journal} {Reviews of Modern Physics}\ }\textbf {\bibinfo {volume}
  {82}},\ \bibinfo {pages} {3045} (\bibinfo {year} {2010})}\BibitemShut
  {NoStop}%
\bibitem [{\citenamefont {Witten}(2016)}]{witten_three_2016}%
  \BibitemOpen
  \bibfield  {author} {\bibinfo {author} {\bibfnamefont {E.}~\bibnamefont
  {Witten}},\ }\href {\doibase 10.1393/ncr/i2016-10125-3} {\bibfield  {journal}
  {\bibinfo  {journal} {La Rivista del Nuovo Cimento}\ }\textbf {\bibinfo
  {volume} {39}},\ \bibinfo {pages} {313} (\bibinfo {year} {2016})},\ \bibinfo
  {note} {arXiv: 1510.07698}\BibitemShut {NoStop}%
\bibitem [{\citenamefont {Schnyder}\ \emph {et~al.}(2008)\citenamefont
  {Schnyder}, \citenamefont {Ryu}, \citenamefont {Furusaki},\ and\
  \citenamefont {Ludwig}}]{schnyder_classification_2008}%
  \BibitemOpen
  \bibfield  {author} {\bibinfo {author} {\bibfnamefont {A.~P.}\ \bibnamefont
  {Schnyder}}, \bibinfo {author} {\bibfnamefont {S.}~\bibnamefont {Ryu}},
  \bibinfo {author} {\bibfnamefont {A.}~\bibnamefont {Furusaki}}, \ and\
  \bibinfo {author} {\bibfnamefont {A.~W.~W.}\ \bibnamefont {Ludwig}},\ }\href
  {\doibase 10.1103/PhysRevB.78.195125} {\bibfield  {journal} {\bibinfo
  {journal} {Physical Review B}\ }\textbf {\bibinfo {volume} {78}},\ \bibinfo
  {pages} {195125} (\bibinfo {year} {2008})}\BibitemShut {NoStop}%
\bibitem [{\citenamefont {Kitaev}(2009)}]{kitaev_periodic_2009}%
  \BibitemOpen
  \bibfield  {author} {\bibinfo {author} {\bibfnamefont {A.}~\bibnamefont
  {Kitaev}},\ }\href {\doibase 10.1063/1.3149495} {\bibfield  {journal}
  {\bibinfo  {journal} {AIP Conference Proceedings}\ }\textbf {\bibinfo
  {volume} {1134}},\ \bibinfo {pages} {22} (\bibinfo {year}
  {2009})}\BibitemShut {NoStop}%
\bibitem [{\citenamefont {Ryu}\ \emph {et~al.}(2010)\citenamefont {Ryu},
  \citenamefont {Schnyder}, \citenamefont {Furusaki},\ and\ \citenamefont
  {Ludwig}}]{ryu_topological_2010}%
  \BibitemOpen
  \bibfield  {author} {\bibinfo {author} {\bibfnamefont {S.}~\bibnamefont
  {Ryu}}, \bibinfo {author} {\bibfnamefont {A.~P.}\ \bibnamefont {Schnyder}},
  \bibinfo {author} {\bibfnamefont {A.}~\bibnamefont {Furusaki}}, \ and\
  \bibinfo {author} {\bibfnamefont {A.~W.~W.}\ \bibnamefont {Ludwig}},\ }\href
  {\doibase 10.1088/1367-2630/12/6/065010} {\bibfield  {journal} {\bibinfo
  {journal} {New Journal of Physics}\ }\textbf {\bibinfo {volume} {12}},\
  \bibinfo {pages} {065010} (\bibinfo {year} {2010})}\BibitemShut {NoStop}%
\bibitem [{\citenamefont {{Nature}}(2016)}]{nature_topological_2016}%
  \BibitemOpen
  \bibfield  {author} {\bibinfo {author} {\bibnamefont {{Nature}}},\ }\href
  {https://www.nature.com/collections/vrlsgqqjbh/} {\emph {\bibinfo {title}
  {Topological semimetals}}}\ (\bibinfo {year} {2016})\BibitemShut {NoStop}%
\bibitem [{\citenamefont {Burkov}(2016)}]{burkov_topological_2016}%
  \BibitemOpen
  \bibfield  {author} {\bibinfo {author} {\bibfnamefont {A.~A.}\ \bibnamefont
  {Burkov}},\ }\href {https://doi.org/10.1038/nmat4788} {\bibfield  {journal}
  {\bibinfo  {journal} {Nature Materials}\ }\textbf {\bibinfo {volume} {15}},\
  \bibinfo {pages} {1145} (\bibinfo {year} {2016})}\BibitemShut {NoStop}%
\bibitem [{\citenamefont {Xu}\ \emph {et~al.}(2015)\citenamefont {Xu},
  \citenamefont {Belopolski}, \citenamefont {Alidoust}, \citenamefont
  {Neupane}, \citenamefont {Bian}, \citenamefont {Zhang}, \citenamefont
  {Sankar}, \citenamefont {Chang}, \citenamefont {Yuan}, \citenamefont {Lee},
  \citenamefont {Huang}, \citenamefont {Zheng}, \citenamefont {Ma},
  \citenamefont {Sanchez}, \citenamefont {Wang}, \citenamefont {Bansil},
  \citenamefont {Chou}, \citenamefont {Shibayev}, \citenamefont {Lin},
  \citenamefont {Jia},\ and\ \citenamefont {Hasan}}]{xu_discovery_2015}%
  \BibitemOpen
  \bibfield  {author} {\bibinfo {author} {\bibfnamefont {S.-Y.}\ \bibnamefont
  {Xu}}, \bibinfo {author} {\bibfnamefont {I.}~\bibnamefont {Belopolski}},
  \bibinfo {author} {\bibfnamefont {N.}~\bibnamefont {Alidoust}}, \bibinfo
  {author} {\bibfnamefont {M.}~\bibnamefont {Neupane}}, \bibinfo {author}
  {\bibfnamefont {G.}~\bibnamefont {Bian}}, \bibinfo {author} {\bibfnamefont
  {C.}~\bibnamefont {Zhang}}, \bibinfo {author} {\bibfnamefont
  {R.}~\bibnamefont {Sankar}}, \bibinfo {author} {\bibfnamefont
  {G.}~\bibnamefont {Chang}}, \bibinfo {author} {\bibfnamefont
  {Z.}~\bibnamefont {Yuan}}, \bibinfo {author} {\bibfnamefont {C.-C.}\
  \bibnamefont {Lee}}, \bibinfo {author} {\bibfnamefont {S.-M.}\ \bibnamefont
  {Huang}}, \bibinfo {author} {\bibfnamefont {H.}~\bibnamefont {Zheng}},
  \bibinfo {author} {\bibfnamefont {J.}~\bibnamefont {Ma}}, \bibinfo {author}
  {\bibfnamefont {D.~S.}\ \bibnamefont {Sanchez}}, \bibinfo {author}
  {\bibfnamefont {B.}~\bibnamefont {Wang}}, \bibinfo {author} {\bibfnamefont
  {A.}~\bibnamefont {Bansil}}, \bibinfo {author} {\bibfnamefont
  {F.}~\bibnamefont {Chou}}, \bibinfo {author} {\bibfnamefont {P.~P.}\
  \bibnamefont {Shibayev}}, \bibinfo {author} {\bibfnamefont {H.}~\bibnamefont
  {Lin}}, \bibinfo {author} {\bibfnamefont {S.}~\bibnamefont {Jia}}, \ and\
  \bibinfo {author} {\bibfnamefont {M.~Z.}\ \bibnamefont {Hasan}},\ }\href
  {\doibase 10.1126/science.aaa9297} {\bibfield  {journal} {\bibinfo  {journal}
  {Science}\ }\textbf {\bibinfo {volume} {349}},\ \bibinfo {pages} {613}
  (\bibinfo {year} {2015})}\BibitemShut {NoStop}%
\bibitem [{\citenamefont {Lu}\ \emph {et~al.}(2015)\citenamefont {Lu},
  \citenamefont {Wang}, \citenamefont {Ye}, \citenamefont {Ran}, \citenamefont
  {Fu}, \citenamefont {Joannopoulos},\ and\ \citenamefont
  {Solja i}}]{lu_experimental_2015}%
  \BibitemOpen
  \bibfield  {author} {\bibinfo {author} {\bibfnamefont {L.}~\bibnamefont
  {Lu}}, \bibinfo {author} {\bibfnamefont {Z.}~\bibnamefont {Wang}}, \bibinfo
  {author} {\bibfnamefont {D.}~\bibnamefont {Ye}}, \bibinfo {author}
  {\bibfnamefont {L.}~\bibnamefont {Ran}}, \bibinfo {author} {\bibfnamefont
  {L.}~\bibnamefont {Fu}}, \bibinfo {author} {\bibfnamefont {J.~D.}\
  \bibnamefont {Joannopoulos}}, \ and\ \bibinfo {author} {\bibfnamefont
  {M.}~\bibnamefont {Solja i}},\ }\href {\doibase 10.1126/science.aaa9273}
  {\bibfield  {journal} {\bibinfo  {journal} {Science}\ }\textbf {\bibinfo
  {volume} {349}},\ \bibinfo {pages} {622} (\bibinfo {year}
  {2015})}\BibitemShut {NoStop}%
\bibitem [{\citenamefont {Lv}\ \emph {et~al.}(2015)\citenamefont {Lv},
  \citenamefont {Weng}, \citenamefont {Fu}, \citenamefont {Wang}, \citenamefont
  {Miao}, \citenamefont {Ma}, \citenamefont {Richard}, \citenamefont {Huang},
  \citenamefont {Zhao}, \citenamefont {Chen}, \citenamefont {Fang},
  \citenamefont {Dai}, \citenamefont {Qian},\ and\ \citenamefont
  {Ding}}]{lv_experimental_2015}%
  \BibitemOpen
  \bibfield  {author} {\bibinfo {author} {\bibfnamefont {B.}~\bibnamefont
  {Lv}}, \bibinfo {author} {\bibfnamefont {H.}~\bibnamefont {Weng}}, \bibinfo
  {author} {\bibfnamefont {B.}~\bibnamefont {Fu}}, \bibinfo {author}
  {\bibfnamefont {X.}~\bibnamefont {Wang}}, \bibinfo {author} {\bibfnamefont
  {H.}~\bibnamefont {Miao}}, \bibinfo {author} {\bibfnamefont {J.}~\bibnamefont
  {Ma}}, \bibinfo {author} {\bibfnamefont {P.}~\bibnamefont {Richard}},
  \bibinfo {author} {\bibfnamefont {X.}~\bibnamefont {Huang}}, \bibinfo
  {author} {\bibfnamefont {L.}~\bibnamefont {Zhao}}, \bibinfo {author}
  {\bibfnamefont {G.}~\bibnamefont {Chen}}, \bibinfo {author} {\bibfnamefont
  {Z.}~\bibnamefont {Fang}}, \bibinfo {author} {\bibfnamefont {X.}~\bibnamefont
  {Dai}}, \bibinfo {author} {\bibfnamefont {T.}~\bibnamefont {Qian}}, \ and\
  \bibinfo {author} {\bibfnamefont {H.}~\bibnamefont {Ding}},\ }\href {\doibase
  10.1103/PhysRevX.5.031013} {\bibfield  {journal} {\bibinfo  {journal}
  {Physical Review X}\ }\textbf {\bibinfo {volume} {5}} (\bibinfo {year}
  {2015}),\ 10.1103/PhysRevX.5.031013}\BibitemShut {NoStop}%
\bibitem [{\citenamefont {Fang}\ \emph {et~al.}(2015)\citenamefont {Fang},
  \citenamefont {Chen}, \citenamefont {Kee},\ and\ \citenamefont
  {Fu}}]{fang_topological_2015}%
  \BibitemOpen
  \bibfield  {author} {\bibinfo {author} {\bibfnamefont {C.}~\bibnamefont
  {Fang}}, \bibinfo {author} {\bibfnamefont {Y.}~\bibnamefont {Chen}}, \bibinfo
  {author} {\bibfnamefont {H.-Y.}\ \bibnamefont {Kee}}, \ and\ \bibinfo
  {author} {\bibfnamefont {L.}~\bibnamefont {Fu}},\ }\href {\doibase
  10.1103/PhysRevB.92.081201} {\bibfield  {journal} {\bibinfo  {journal}
  {Physical Review B}\ }\textbf {\bibinfo {volume} {92}},\ \bibinfo {pages}
  {081201} (\bibinfo {year} {2015})}\BibitemShut {NoStop}%
\bibitem [{\citenamefont {Wang}\ and\ \citenamefont
  {Chou}(1995)}]{wang_structural_1995}%
  \BibitemOpen
  \bibfield  {author} {\bibinfo {author} {\bibfnamefont {Y.}~\bibnamefont
  {Wang}}\ and\ \bibinfo {author} {\bibfnamefont {M.~Y.}\ \bibnamefont
  {Chou}},\ }\href {\doibase 10.1103/PhysRevB.51.7500} {\bibfield  {journal}
  {\bibinfo  {journal} {Physical Review B}\ }\textbf {\bibinfo {volume} {51}},\
  \bibinfo {pages} {7500} (\bibinfo {year} {1995})}\BibitemShut {NoStop}%
\bibitem [{\citenamefont {Ahuja}\ \emph {et~al.}(1997)\citenamefont {Ahuja},
  \citenamefont {Johansson}, \citenamefont {Wills},\ and\ \citenamefont
  {Eriksson}}]{ahuja_semiconducting_1997}%
  \BibitemOpen
  \bibfield  {author} {\bibinfo {author} {\bibfnamefont {R.}~\bibnamefont
  {Ahuja}}, \bibinfo {author} {\bibfnamefont {B.}~\bibnamefont {Johansson}},
  \bibinfo {author} {\bibfnamefont {J.~M.}\ \bibnamefont {Wills}}, \ and\
  \bibinfo {author} {\bibfnamefont {O.}~\bibnamefont {Eriksson}},\ }\href
  {\doibase 10.1063/1.120371} {\bibfield  {journal} {\bibinfo  {journal}
  {Applied Physics Letters}\ }\textbf {\bibinfo {volume} {71}},\ \bibinfo
  {pages} {3498} (\bibinfo {year} {1997})}\BibitemShut {NoStop}%
\bibitem [{\citenamefont {Palasyuk}\ and\ \citenamefont
  {Tkacz}(2004)}]{palasyuk_pressure_2004}%
  \BibitemOpen
  \bibfield  {author} {\bibinfo {author} {\bibfnamefont {T.}~\bibnamefont
  {Palasyuk}}\ and\ \bibinfo {author} {\bibfnamefont {M.}~\bibnamefont
  {Tkacz}},\ }\href {\doibase 10.1016/j.ssc.2004.01.040} {\bibfield  {journal}
  {\bibinfo  {journal} {Solid State Communications}\ }\textbf {\bibinfo
  {volume} {130}},\ \bibinfo {pages} {219} (\bibinfo {year}
  {2004})}\BibitemShut {NoStop}%
\bibitem [{\citenamefont {Ohmura}\ \emph {et~al.}(2006)\citenamefont {Ohmura},
  \citenamefont {Machida}, \citenamefont {Watanuki}, \citenamefont {Aoki},
  \citenamefont {Nakano},\ and\ \citenamefont
  {Takemura}}]{ohmura_infrared_2006}%
  \BibitemOpen
  \bibfield  {author} {\bibinfo {author} {\bibfnamefont {A.}~\bibnamefont
  {Ohmura}}, \bibinfo {author} {\bibfnamefont {A.}~\bibnamefont {Machida}},
  \bibinfo {author} {\bibfnamefont {T.}~\bibnamefont {Watanuki}}, \bibinfo
  {author} {\bibfnamefont {K.}~\bibnamefont {Aoki}}, \bibinfo {author}
  {\bibfnamefont {S.}~\bibnamefont {Nakano}}, \ and\ \bibinfo {author}
  {\bibfnamefont {K.}~\bibnamefont {Takemura}},\ }\href {\doibase
  10.1103/PhysRevB.73.104105} {\bibfield  {journal} {\bibinfo  {journal}
  {Physical Review B}\ }\textbf {\bibinfo {volume} {73}},\ \bibinfo {pages}
  {104105} (\bibinfo {year} {2006})}\BibitemShut {NoStop}%
\bibitem [{\citenamefont {Machida}\ \emph {et~al.}(2006)\citenamefont
  {Machida}, \citenamefont {Ohmura}, \citenamefont {Watanuki}, \citenamefont
  {Ikeda}, \citenamefont {Aoki}, \citenamefont {Nakano},\ and\ \citenamefont
  {Takemura}}]{machida_x-ray_2006}%
  \BibitemOpen
  \bibfield  {author} {\bibinfo {author} {\bibfnamefont {A.}~\bibnamefont
  {Machida}}, \bibinfo {author} {\bibfnamefont {A.}~\bibnamefont {Ohmura}},
  \bibinfo {author} {\bibfnamefont {T.}~\bibnamefont {Watanuki}}, \bibinfo
  {author} {\bibfnamefont {T.}~\bibnamefont {Ikeda}}, \bibinfo {author}
  {\bibfnamefont {K.}~\bibnamefont {Aoki}}, \bibinfo {author} {\bibfnamefont
  {S.}~\bibnamefont {Nakano}}, \ and\ \bibinfo {author} {\bibfnamefont
  {K.}~\bibnamefont {Takemura}},\ }\href {\doibase 10.1016/j.ssc.2006.04.011}
  {\bibfield  {journal} {\bibinfo  {journal} {Solid State Communications}\
  }\textbf {\bibinfo {volume} {138}},\ \bibinfo {pages} {436} (\bibinfo {year}
  {2006})}\BibitemShut {NoStop}%
\bibitem [{\citenamefont {Remhof}\ \emph {et~al.}(1999)\citenamefont {Remhof},
  \citenamefont {Song}, \citenamefont {Sutter}, \citenamefont {Schreyer},
  \citenamefont {Siebrecht}, \citenamefont {Zabel}, \citenamefont {Guthoff},\
  and\ \citenamefont {Windgasse}}]{remhof_hydrogen_1999}%
  \BibitemOpen
  \bibfield  {author} {\bibinfo {author} {\bibfnamefont {A.}~\bibnamefont
  {Remhof}}, \bibinfo {author} {\bibfnamefont {G.}~\bibnamefont {Song}},
  \bibinfo {author} {\bibfnamefont {C.}~\bibnamefont {Sutter}}, \bibinfo
  {author} {\bibfnamefont {A.}~\bibnamefont {Schreyer}}, \bibinfo {author}
  {\bibfnamefont {R.}~\bibnamefont {Siebrecht}}, \bibinfo {author}
  {\bibfnamefont {H.}~\bibnamefont {Zabel}}, \bibinfo {author} {\bibfnamefont
  {F.}~\bibnamefont {Guthoff}}, \ and\ \bibinfo {author} {\bibfnamefont
  {J.}~\bibnamefont {Windgasse}},\ }\href {\doibase 10.1103/PhysRevB.59.6689}
  {\bibfield  {journal} {\bibinfo  {journal} {Physical Review B}\ }\textbf
  {\bibinfo {volume} {59}},\ \bibinfo {pages} {6689} (\bibinfo {year}
  {1999})}\BibitemShut {NoStop}%
\bibitem [{\citenamefont {Wolf}\ and\ \citenamefont
  {Herzig}(2002)}]{wolf_first-principles_2002}%
  \BibitemOpen
  \bibfield  {author} {\bibinfo {author} {\bibfnamefont {W.}~\bibnamefont
  {Wolf}}\ and\ \bibinfo {author} {\bibfnamefont {P.}~\bibnamefont {Herzig}},\
  }\href {\doibase 10.1103/PhysRevB.66.224112} {\bibfield  {journal} {\bibinfo
  {journal} {Physical Review B}\ }\textbf {\bibinfo {volume} {66}},\ \bibinfo
  {pages} {224112} (\bibinfo {year} {2002})}\BibitemShut {NoStop}%
\bibitem [{\citenamefont {de~Almeida}\ \emph {et~al.}(2009)\citenamefont
  {de~Almeida}, \citenamefont {Kim}, \citenamefont {Ortiz}, \citenamefont
  {Klintenberg},\ and\ \citenamefont {Ahuja}}]{de_almeida_dynamical_2009}%
  \BibitemOpen
  \bibfield  {author} {\bibinfo {author} {\bibfnamefont {J.~S.}\ \bibnamefont
  {de~Almeida}}, \bibinfo {author} {\bibfnamefont {D.~Y.}\ \bibnamefont {Kim}},
  \bibinfo {author} {\bibfnamefont {C.}~\bibnamefont {Ortiz}}, \bibinfo
  {author} {\bibfnamefont {M.}~\bibnamefont {Klintenberg}}, \ and\ \bibinfo
  {author} {\bibfnamefont {R.}~\bibnamefont {Ahuja}},\ }\href {\doibase
  10.1063/1.3155505} {\bibfield  {journal} {\bibinfo  {journal} {Applied
  Physics Letters}\ }\textbf {\bibinfo {volume} {94}},\ \bibinfo {pages}
  {251913} (\bibinfo {year} {2009})}\BibitemShut {NoStop}%
\bibitem [{\citenamefont {Shao}\ \emph {et~al.}(2018)\citenamefont {Shao},
  \citenamefont {Chen}, \citenamefont {Gu}, \citenamefont {Guo}, \citenamefont
  {Lu}, \citenamefont {Sun}, \citenamefont {Sheng},\ and\ \citenamefont
  {Xing}}]{shao_nonsymmorphic_2018}%
  \BibitemOpen
  \bibfield  {author} {\bibinfo {author} {\bibfnamefont {D.}~\bibnamefont
  {Shao}}, \bibinfo {author} {\bibfnamefont {T.}~\bibnamefont {Chen}}, \bibinfo
  {author} {\bibfnamefont {Q.}~\bibnamefont {Gu}}, \bibinfo {author}
  {\bibfnamefont {Z.}~\bibnamefont {Guo}}, \bibinfo {author} {\bibfnamefont
  {P.}~\bibnamefont {Lu}}, \bibinfo {author} {\bibfnamefont {J.}~\bibnamefont
  {Sun}}, \bibinfo {author} {\bibfnamefont {L.}~\bibnamefont {Sheng}}, \ and\
  \bibinfo {author} {\bibfnamefont {D.}~\bibnamefont {Xing}},\ }\href {\doibase
  10.1038/s41598-018-19870-5} {\bibfield  {journal} {\bibinfo  {journal}
  {Scientific Reports}\ }\textbf {\bibinfo {volume} {8}} (\bibinfo {year}
  {2018}),\ 10.1038/s41598-018-19870-5}\BibitemShut {NoStop}%
\bibitem [{\citenamefont {Wang}\ \emph {et~al.}(2018)\citenamefont {Wang},
  \citenamefont {Liu}, \citenamefont {Jin}, \citenamefont {Sui}, \citenamefont
  {Zhang}, \citenamefont {Duan}, \citenamefont {Liu},\ and\ \citenamefont
  {Huang}}]{wang_pseudo_2018}%
  \BibitemOpen
  \bibfield  {author} {\bibinfo {author} {\bibfnamefont {J.}~\bibnamefont
  {Wang}}, \bibinfo {author} {\bibfnamefont {Y.}~\bibnamefont {Liu}}, \bibinfo
  {author} {\bibfnamefont {K.-H.}\ \bibnamefont {Jin}}, \bibinfo {author}
  {\bibfnamefont {X.}~\bibnamefont {Sui}}, \bibinfo {author} {\bibfnamefont
  {L.}~\bibnamefont {Zhang}}, \bibinfo {author} {\bibfnamefont
  {W.}~\bibnamefont {Duan}}, \bibinfo {author} {\bibfnamefont {F.}~\bibnamefont
  {Liu}}, \ and\ \bibinfo {author} {\bibfnamefont {B.}~\bibnamefont {Huang}},\
  }\href {\doibase 10.1103/PhysRevB.98.201112} {\bibfield  {journal} {\bibinfo
  {journal} {Physical Review B}\ }\textbf {\bibinfo {volume} {98}} (\bibinfo
  {year} {2018}),\ 10.1103/PhysRevB.98.201112}\BibitemShut {NoStop}%
\bibitem [{\citenamefont {Persson}(2016)}]{persson_materials_2016-1}%
  \BibitemOpen
  \bibfield  {author} {\bibinfo {author} {\bibfnamefont {K.}~\bibnamefont
  {Persson}},\ }\href {\doibase 10.17188/1199674} {\emph {\bibinfo {title}
  {Materials {Data} on {YH}3 ({SG}:165) by {Materials} {Project}}}}\ (\bibinfo
  {year} {2016})\BibitemShut {NoStop}%
\bibitem [{\citenamefont {I.~Aroyo}\ \emph {et~al.}(2011)\citenamefont
  {I.~Aroyo}, \citenamefont {M.~Perez-Mato}, \citenamefont {Orobengoa},
  \citenamefont {Tasci}, \citenamefont {De~la Flor~Martin},\ and\ \citenamefont
  {Kirov}}]{i._aroyo_crystallography_2011}%
  \BibitemOpen
  \bibfield  {author} {\bibinfo {author} {\bibfnamefont {M.}~\bibnamefont
  {I.~Aroyo}}, \bibinfo {author} {\bibfnamefont {J.}~\bibnamefont
  {M.~Perez-Mato}}, \bibinfo {author} {\bibfnamefont {D.}~\bibnamefont
  {Orobengoa}}, \bibinfo {author} {\bibfnamefont {E.}~\bibnamefont {Tasci}},
  \bibinfo {author} {\bibfnamefont {G.}~\bibnamefont {De~la Flor~Martin}}, \
  and\ \bibinfo {author} {\bibfnamefont {A.}~\bibnamefont {Kirov}},\
  }\href@noop {} {\bibfield  {journal} {\bibinfo  {journal} {Bulgarian Chemical
  Communications}\ }\textbf {\bibinfo {volume} {43}},\ \bibinfo {pages} {183}
  (\bibinfo {year} {2011})}\BibitemShut {NoStop}%
\bibitem [{\citenamefont {Aroyo}\ \emph
  {et~al.}(2006{\natexlab{a}})\citenamefont {Aroyo}, \citenamefont
  {Perez-Mato}, \citenamefont {Capillas}, \citenamefont {Kroumova},
  \citenamefont {Ivantchev}, \citenamefont {Madariaga}, \citenamefont {Kirov},\
  and\ \citenamefont {Wondratschek}}]{aroyo_bilbao_2006}%
  \BibitemOpen
  \bibfield  {author} {\bibinfo {author} {\bibfnamefont {M.~I.}\ \bibnamefont
  {Aroyo}}, \bibinfo {author} {\bibfnamefont {J.~M.}\ \bibnamefont
  {Perez-Mato}}, \bibinfo {author} {\bibfnamefont {C.}~\bibnamefont
  {Capillas}}, \bibinfo {author} {\bibfnamefont {E.}~\bibnamefont {Kroumova}},
  \bibinfo {author} {\bibfnamefont {S.}~\bibnamefont {Ivantchev}}, \bibinfo
  {author} {\bibfnamefont {G.}~\bibnamefont {Madariaga}}, \bibinfo {author}
  {\bibfnamefont {A.}~\bibnamefont {Kirov}}, \ and\ \bibinfo {author}
  {\bibfnamefont {H.}~\bibnamefont {Wondratschek}},\ }\href {\doibase
  10.1524/zkri.2006.221.1.15} {\bibfield  {journal} {\bibinfo  {journal}
  {Zeitschrift fur Kristallographie - Crystalline Materials}\ }\textbf
  {\bibinfo {volume} {221}} (\bibinfo {year} {2006}{\natexlab{a}}),\
  10.1524/zkri.2006.221.1.15}\BibitemShut {NoStop}%
\bibitem [{\citenamefont {Aroyo}\ \emph
  {et~al.}(2006{\natexlab{b}})\citenamefont {Aroyo}, \citenamefont {Kirov},
  \citenamefont {Capillas}, \citenamefont {Perez-Mato},\ and\ \citenamefont
  {Wondratschek}}]{aroyo_bilbao_2006-1}%
  \BibitemOpen
  \bibfield  {author} {\bibinfo {author} {\bibfnamefont {M.~I.}\ \bibnamefont
  {Aroyo}}, \bibinfo {author} {\bibfnamefont {A.}~\bibnamefont {Kirov}},
  \bibinfo {author} {\bibfnamefont {C.}~\bibnamefont {Capillas}}, \bibinfo
  {author} {\bibfnamefont {J.~M.}\ \bibnamefont {Perez-Mato}}, \ and\ \bibinfo
  {author} {\bibfnamefont {H.}~\bibnamefont {Wondratschek}},\ }\href {\doibase
  10.1107/S0108767305040286} {\bibfield  {journal} {\bibinfo  {journal} {Acta
  Crystallographica Section A Foundations of Crystallography}\ }\textbf
  {\bibinfo {volume} {62}},\ \bibinfo {pages} {115} (\bibinfo {year}
  {2006}{\natexlab{b}})}\BibitemShut {NoStop}%
\bibitem [{\citenamefont {Hohenberg}\ and\ \citenamefont
  {Kohn}(1964)}]{hohenberg_inhomogeneous_1964}%
  \BibitemOpen
  \bibfield  {author} {\bibinfo {author} {\bibfnamefont {P.}~\bibnamefont
  {Hohenberg}}\ and\ \bibinfo {author} {\bibfnamefont {W.}~\bibnamefont
  {Kohn}},\ }\href {\doibase 10.1103/PhysRev.136.B864} {\bibfield  {journal}
  {\bibinfo  {journal} {Physical Review}\ }\textbf {\bibinfo {volume} {136}},\
  \bibinfo {pages} {B864} (\bibinfo {year} {1964})}\BibitemShut {NoStop}%
\bibitem [{\citenamefont {Kohn}\ and\ \citenamefont
  {Sham}(1965)}]{kohn_self-consistent_1965}%
  \BibitemOpen
  \bibfield  {author} {\bibinfo {author} {\bibfnamefont {W.}~\bibnamefont
  {Kohn}}\ and\ \bibinfo {author} {\bibfnamefont {L.~J.}\ \bibnamefont
  {Sham}},\ }\href {\doibase 10.1103/PhysRev.140.A1133} {\bibfield  {journal}
  {\bibinfo  {journal} {Physical Review}\ }\textbf {\bibinfo {volume} {140}},\
  \bibinfo {pages} {A1133} (\bibinfo {year} {1965})}\BibitemShut {NoStop}%
\bibitem [{\citenamefont {Kresse}\ and\ \citenamefont
  {Furthmuller}(1996{\natexlab{a}})}]{kresse_efficient_1996}%
  \BibitemOpen
  \bibfield  {author} {\bibinfo {author} {\bibfnamefont {G.}~\bibnamefont
  {Kresse}}\ and\ \bibinfo {author} {\bibfnamefont {J.}~\bibnamefont
  {Furthmuller}},\ }\href {\doibase 10.1103/PhysRevB.54.11169} {\bibfield
  {journal} {\bibinfo  {journal} {Physical Review B}\ }\textbf {\bibinfo
  {volume} {54}},\ \bibinfo {pages} {11169} (\bibinfo {year}
  {1996}{\natexlab{a}})}\BibitemShut {NoStop}%
\bibitem [{\citenamefont {Kresse}\ and\ \citenamefont
  {Furthmuller}(1996{\natexlab{b}})}]{kresse_efficiency_1996}%
  \BibitemOpen
  \bibfield  {author} {\bibinfo {author} {\bibfnamefont {G.}~\bibnamefont
  {Kresse}}\ and\ \bibinfo {author} {\bibfnamefont {J.}~\bibnamefont
  {Furthmuller}},\ }\href {\doibase 10.1016/0927-0256(96)00008-0} {\bibfield
  {journal} {\bibinfo  {journal} {Computational Materials Science}\ }\textbf
  {\bibinfo {volume} {6}},\ \bibinfo {pages} {15} (\bibinfo {year}
  {1996}{\natexlab{b}})}\BibitemShut {NoStop}%
\bibitem [{\citenamefont {Blochl}(1994)}]{blochl_projector_1994}%
  \BibitemOpen
  \bibfield  {author} {\bibinfo {author} {\bibfnamefont {P.~E.}\ \bibnamefont
  {Blochl}},\ }\href {\doibase 10.1103/PhysRevB.50.17953} {\bibfield  {journal}
  {\bibinfo  {journal} {Physical Review B}\ }\textbf {\bibinfo {volume} {50}},\
  \bibinfo {pages} {17953} (\bibinfo {year} {1994})}\BibitemShut {NoStop}%
\bibitem [{\citenamefont {Kresse}\ and\ \citenamefont
  {Joubert}(1999)}]{kresse_ultrasoft_1999}%
  \BibitemOpen
  \bibfield  {author} {\bibinfo {author} {\bibfnamefont {G.}~\bibnamefont
  {Kresse}}\ and\ \bibinfo {author} {\bibfnamefont {D.}~\bibnamefont
  {Joubert}},\ }\href {\doibase 10.1103/PhysRevB.59.1758} {\bibfield  {journal}
  {\bibinfo  {journal} {Physical Review B}\ }\textbf {\bibinfo {volume} {59}},\
  \bibinfo {pages} {1758} (\bibinfo {year} {1999})}\BibitemShut {NoStop}%
\bibitem [{\citenamefont {Perdew}\ \emph {et~al.}(1996)\citenamefont {Perdew},
  \citenamefont {Burke},\ and\ \citenamefont
  {Ernzerhof}}]{perdew_generalized_1996}%
  \BibitemOpen
  \bibfield  {author} {\bibinfo {author} {\bibfnamefont {J.~P.}\ \bibnamefont
  {Perdew}}, \bibinfo {author} {\bibfnamefont {K.}~\bibnamefont {Burke}}, \
  and\ \bibinfo {author} {\bibfnamefont {M.}~\bibnamefont {Ernzerhof}},\ }\href
  {\doibase 10.1103/PhysRevLett.77.3865} {\bibfield  {journal} {\bibinfo
  {journal} {Physical Review Letters}\ }\textbf {\bibinfo {volume} {77}},\
  \bibinfo {pages} {3865} (\bibinfo {year} {1996})}\BibitemShut {NoStop}%
\bibitem [{\citenamefont {Monkhorst}\ and\ \citenamefont
  {Pack}(1976)}]{monkhorst_special_1976}%
  \BibitemOpen
  \bibfield  {author} {\bibinfo {author} {\bibfnamefont {H.~J.}\ \bibnamefont
  {Monkhorst}}\ and\ \bibinfo {author} {\bibfnamefont {J.~D.}\ \bibnamefont
  {Pack}},\ }\href {\doibase 10.1103/PhysRevB.13.5188} {\bibfield  {journal}
  {\bibinfo  {journal} {Physical Review B}\ }\textbf {\bibinfo {volume} {13}},\
  \bibinfo {pages} {5188} (\bibinfo {year} {1976})}\BibitemShut {NoStop}%
\bibitem [{\citenamefont {Kruthoff}\ \emph {et~al.}(2017)\citenamefont
  {Kruthoff}, \citenamefont {de~Boer}, \citenamefont {van Wezel}, \citenamefont
  {Kane},\ and\ \citenamefont {Slager}}]{kruthoff_topological_2017}%
  \BibitemOpen
  \bibfield  {author} {\bibinfo {author} {\bibfnamefont {J.}~\bibnamefont
  {Kruthoff}}, \bibinfo {author} {\bibfnamefont {J.}~\bibnamefont {de~Boer}},
  \bibinfo {author} {\bibfnamefont {J.}~\bibnamefont {van Wezel}}, \bibinfo
  {author} {\bibfnamefont {C.~L.}\ \bibnamefont {Kane}}, \ and\ \bibinfo
  {author} {\bibfnamefont {R.-J.}\ \bibnamefont {Slager}},\ }\href {\doibase
  10.1103/PhysRevX.7.041069} {\bibfield  {journal} {\bibinfo  {journal}
  {Physical Review X}\ }\textbf {\bibinfo {volume} {7}},\ \bibinfo {pages}
  {041069} (\bibinfo {year} {2017})}\BibitemShut {NoStop}%
\bibitem [{\citenamefont {Bradlyn}\ \emph {et~al.}(2017)\citenamefont
  {Bradlyn}, \citenamefont {Elcoro}, \citenamefont {Cano}, \citenamefont
  {Vergniory}, \citenamefont {Wang}, \citenamefont {Felser}, \citenamefont
  {Aroyo},\ and\ \citenamefont {Bernevig}}]{bradlyn_topological_2017}%
  \BibitemOpen
  \bibfield  {author} {\bibinfo {author} {\bibfnamefont {B.}~\bibnamefont
  {Bradlyn}}, \bibinfo {author} {\bibfnamefont {L.}~\bibnamefont {Elcoro}},
  \bibinfo {author} {\bibfnamefont {J.}~\bibnamefont {Cano}}, \bibinfo {author}
  {\bibfnamefont {M.~G.}\ \bibnamefont {Vergniory}}, \bibinfo {author}
  {\bibfnamefont {Z.}~\bibnamefont {Wang}}, \bibinfo {author} {\bibfnamefont
  {C.}~\bibnamefont {Felser}}, \bibinfo {author} {\bibfnamefont {M.~I.}\
  \bibnamefont {Aroyo}}, \ and\ \bibinfo {author} {\bibfnamefont {B.~A.}\
  \bibnamefont {Bernevig}},\ }\href {\doibase 10.1038/nature23268} {\bibfield
  {journal} {\bibinfo  {journal} {Nature}\ }\textbf {\bibinfo {volume} {547}},\
  \bibinfo {pages} {298} (\bibinfo {year} {2017})},\ \bibinfo {note} {arXiv:
  1703.02050}\BibitemShut {NoStop}%
\bibitem [{\citenamefont {Vergniory}\ \emph {et~al.}(2017)\citenamefont
  {Vergniory}, \citenamefont {Elcoro}, \citenamefont {Wang}, \citenamefont
  {Cano}, \citenamefont {Felser}, \citenamefont {Aroyo}, \citenamefont
  {Bernevig},\ and\ \citenamefont {Bradlyn}}]{vergniory_graph_2017}%
  \BibitemOpen
  \bibfield  {author} {\bibinfo {author} {\bibfnamefont {M.~G.}\ \bibnamefont
  {Vergniory}}, \bibinfo {author} {\bibfnamefont {L.}~\bibnamefont {Elcoro}},
  \bibinfo {author} {\bibfnamefont {Z.}~\bibnamefont {Wang}}, \bibinfo {author}
  {\bibfnamefont {J.}~\bibnamefont {Cano}}, \bibinfo {author} {\bibfnamefont
  {C.}~\bibnamefont {Felser}}, \bibinfo {author} {\bibfnamefont {M.~I.}\
  \bibnamefont {Aroyo}}, \bibinfo {author} {\bibfnamefont {B.~A.}\ \bibnamefont
  {Bernevig}}, \ and\ \bibinfo {author} {\bibfnamefont {B.}~\bibnamefont
  {Bradlyn}},\ }\href {\doibase 10.1103/PhysRevE.96.023310} {\bibfield
  {journal} {\bibinfo  {journal} {Physical Review E}\ }\textbf {\bibinfo
  {volume} {96}},\ \bibinfo {pages} {023310} (\bibinfo {year} {2017})},\
  \bibinfo {note} {arXiv: 1706.08529}\BibitemShut {NoStop}%
\bibitem [{\citenamefont {Bradlyn}\ \emph {et~al.}(2018)\citenamefont
  {Bradlyn}, \citenamefont {Elcoro}, \citenamefont {Vergniory}, \citenamefont
  {Cano}, \citenamefont {Wang}, \citenamefont {Felser}, \citenamefont {Aroyo},\
  and\ \citenamefont {Bernevig}}]{bradlyn_band_2018}%
  \BibitemOpen
  \bibfield  {author} {\bibinfo {author} {\bibfnamefont {B.}~\bibnamefont
  {Bradlyn}}, \bibinfo {author} {\bibfnamefont {L.}~\bibnamefont {Elcoro}},
  \bibinfo {author} {\bibfnamefont {M.~G.}\ \bibnamefont {Vergniory}}, \bibinfo
  {author} {\bibfnamefont {J.}~\bibnamefont {Cano}}, \bibinfo {author}
  {\bibfnamefont {Z.}~\bibnamefont {Wang}}, \bibinfo {author} {\bibfnamefont
  {C.}~\bibnamefont {Felser}}, \bibinfo {author} {\bibfnamefont {M.~I.}\
  \bibnamefont {Aroyo}}, \ and\ \bibinfo {author} {\bibfnamefont {B.~A.}\
  \bibnamefont {Bernevig}},\ }\href {\doibase 10.1103/PhysRevB.97.035138}
  {\bibfield  {journal} {\bibinfo  {journal} {Physical Review B}\ }\textbf
  {\bibinfo {volume} {97}},\ \bibinfo {pages} {035138} (\bibinfo {year}
  {2018})}\BibitemShut {NoStop}%
\bibitem [{\citenamefont {Mostofi}\ \emph {et~al.}(2014)\citenamefont
  {Mostofi}, \citenamefont {Yates}, \citenamefont {Pizzi}, \citenamefont {Lee},
  \citenamefont {Souza}, \citenamefont {Vanderbilt},\ and\ \citenamefont
  {Marzari}}]{mostofi_updated_2014}%
  \BibitemOpen
  \bibfield  {author} {\bibinfo {author} {\bibfnamefont {A.~A.}\ \bibnamefont
  {Mostofi}}, \bibinfo {author} {\bibfnamefont {J.~R.}\ \bibnamefont {Yates}},
  \bibinfo {author} {\bibfnamefont {G.}~\bibnamefont {Pizzi}}, \bibinfo
  {author} {\bibfnamefont {Y.-S.}\ \bibnamefont {Lee}}, \bibinfo {author}
  {\bibfnamefont {I.}~\bibnamefont {Souza}}, \bibinfo {author} {\bibfnamefont
  {D.}~\bibnamefont {Vanderbilt}}, \ and\ \bibinfo {author} {\bibfnamefont
  {N.}~\bibnamefont {Marzari}},\ }\href {\doibase 10.1016/j.cpc.2014.05.003}
  {\bibfield  {journal} {\bibinfo  {journal} {Computer Physics Communications}\
  }\textbf {\bibinfo {volume} {185}},\ \bibinfo {pages} {2309} (\bibinfo {year}
  {2014})}\BibitemShut {NoStop}%
\bibitem [{\citenamefont {Wu}\ \emph {et~al.}(2018)\citenamefont {Wu},
  \citenamefont {Zhang}, \citenamefont {Song}, \citenamefont {Troyer},\ and\
  \citenamefont {Soluyanov}}]{wu_wanniertools_2018}%
  \BibitemOpen
  \bibfield  {author} {\bibinfo {author} {\bibfnamefont {Q.}~\bibnamefont
  {Wu}}, \bibinfo {author} {\bibfnamefont {S.}~\bibnamefont {Zhang}}, \bibinfo
  {author} {\bibfnamefont {H.-F.}\ \bibnamefont {Song}}, \bibinfo {author}
  {\bibfnamefont {M.}~\bibnamefont {Troyer}}, \ and\ \bibinfo {author}
  {\bibfnamefont {A.~A.}\ \bibnamefont {Soluyanov}},\ }\href {\doibase
  https://doi.org/10.1016/j.cpc.2017.09.033} {\bibfield  {journal} {\bibinfo
  {journal} {Computer Physics Communications}\ }\textbf {\bibinfo {volume}
  {224}},\ \bibinfo {pages} {405 } (\bibinfo {year} {2018})}\BibitemShut
  {NoStop}%
\bibitem [{\citenamefont {Ashcroft}(2004)}]{ashcroft_hydrogen_2004}%
  \BibitemOpen
  \bibfield  {author} {\bibinfo {author} {\bibfnamefont {N.~W.}\ \bibnamefont
  {Ashcroft}},\ }\href {\doibase 10.1103/PhysRevLett.92.187002} {\bibfield
  {journal} {\bibinfo  {journal} {Physical Review Letters}\ }\textbf {\bibinfo
  {volume} {92}},\ \bibinfo {pages} {187002} (\bibinfo {year}
  {2004})}\BibitemShut {NoStop}%
\bibitem [{\citenamefont {Liu}\ \emph {et~al.}(2017{\natexlab{a}})\citenamefont
  {Liu}, \citenamefont {Sun}, \citenamefont {Wang},\ and\ \citenamefont
  {Lu}}]{liu_high-pressure_2017}%
  \BibitemOpen
  \bibfield  {author} {\bibinfo {author} {\bibfnamefont {L.-L.}\ \bibnamefont
  {Liu}}, \bibinfo {author} {\bibfnamefont {H.-J.}\ \bibnamefont {Sun}},
  \bibinfo {author} {\bibfnamefont {C.~Z.}\ \bibnamefont {Wang}}, \ and\
  \bibinfo {author} {\bibfnamefont {W.-C.}\ \bibnamefont {Lu}},\ }\href
  {\doibase 10.1088/1361-648x/aa787d} {\bibfield  {journal} {\bibinfo
  {journal} {Journal of Physics: Condensed Matter}\ }\textbf {\bibinfo {volume}
  {29}},\ \bibinfo {pages} {325401} (\bibinfo {year}
  {2017}{\natexlab{a}})}\BibitemShut {NoStop}%
\bibitem [{\citenamefont {Liu}\ \emph {et~al.}(2017{\natexlab{b}})\citenamefont
  {Liu}, \citenamefont {Naumov}, \citenamefont {Hoffmann}, \citenamefont
  {Ashcroft},\ and\ \citenamefont {Hemley}}]{liu_potential_2017}%
  \BibitemOpen
  \bibfield  {author} {\bibinfo {author} {\bibfnamefont {H.}~\bibnamefont
  {Liu}}, \bibinfo {author} {\bibfnamefont {I.~I.}\ \bibnamefont {Naumov}},
  \bibinfo {author} {\bibfnamefont {R.}~\bibnamefont {Hoffmann}}, \bibinfo
  {author} {\bibfnamefont {N.~W.}\ \bibnamefont {Ashcroft}}, \ and\ \bibinfo
  {author} {\bibfnamefont {R.~J.}\ \bibnamefont {Hemley}},\ }\href {\doibase
  10.1073/pnas.1704505114} {\bibfield  {journal} {\bibinfo  {journal}
  {Proceedings of the National Academy of Sciences}\ }\textbf {\bibinfo
  {volume} {114}},\ \bibinfo {pages} {6990} (\bibinfo {year}
  {2017}{\natexlab{b}})}\BibitemShut {NoStop}%
\bibitem [{\citenamefont {Peng}\ \emph {et~al.}(2017)\citenamefont {Peng},
  \citenamefont {Sun}, \citenamefont {Pickard}, \citenamefont {Needs},
  \citenamefont {Wu},\ and\ \citenamefont {Ma}}]{peng_hydrogen_2017}%
  \BibitemOpen
  \bibfield  {author} {\bibinfo {author} {\bibfnamefont {F.}~\bibnamefont
  {Peng}}, \bibinfo {author} {\bibfnamefont {Y.}~\bibnamefont {Sun}}, \bibinfo
  {author} {\bibfnamefont {C.~J.}\ \bibnamefont {Pickard}}, \bibinfo {author}
  {\bibfnamefont {R.~J.}\ \bibnamefont {Needs}}, \bibinfo {author}
  {\bibfnamefont {Q.}~\bibnamefont {Wu}}, \ and\ \bibinfo {author}
  {\bibfnamefont {Y.}~\bibnamefont {Ma}},\ }\href {\doibase
  10.1103/PhysRevLett.119.107001} {\bibfield  {journal} {\bibinfo  {journal}
  {Physical Review Letters}\ }\textbf {\bibinfo {volume} {119}},\ \bibinfo
  {pages} {107001} (\bibinfo {year} {2017})}\BibitemShut {NoStop}%
\bibitem [{\citenamefont {Keyes}(1960)}]{keyes_effects_1960}%
  \BibitemOpen
  \bibfield  {author} {\bibinfo {author} {\bibfnamefont {R.~W.}\ \bibnamefont
  {Keyes}},\ }in\ \href {\doibase 10.1016/S0081-1947(08)60168-X} {\emph
  {\bibinfo {booktitle} {Solid {State} {Physics}}}},\ Vol.~\bibinfo {volume}
  {11},\ \bibinfo {editor} {edited by\ \bibinfo {editor} {\bibfnamefont
  {F.}~\bibnamefont {SEITZ}}\ and\ \bibinfo {editor} {\bibfnamefont
  {D.}~\bibnamefont {TURNBULL}}}\ (\bibinfo  {publisher} {Academic Press},\
  \bibinfo {year} {1960})\ pp.\ \bibinfo {pages} {149--221}\BibitemShut
  {NoStop}%
\end{thebibliography}%

\end{document}